\documentclass{article}

\usepackage{arxiv}
\usepackage{eurosym}
\usepackage[utf8]{inputenc} 
\usepackage[T1]{fontenc}    
\usepackage{hyperref}       
\usepackage{url}            
\usepackage{booktabs}       
\usepackage{amsfonts}       
\usepackage{nicefrac}       
\usepackage{microtype}      
\usepackage{lipsum}
\usepackage{graphicx}
\usepackage{color}
\usepackage{amssymb}
\usepackage[ruled,vlined]{algorithm2e}
\usepackage{subcaption}
\usepackage{amsmath}
\usepackage[autostyle]{csquotes}

\title{Data fusion strategies for energy efficiency in buildings: Overview, challenges and novel orientations}

\author{
  Yassine Himeur\thanks{This paper has been accepted in Information Fusion } , Abdullah Alsalemi, Ayman Al-Kababji, Faycal Bensaali\\
  Department of Electrical Engineering\\
  Qatar University\\
  Doha, Qatar \\
  \texttt{yassine.himeur@qu.edu.qa;a.alsalemi@qu.edu.qa;aa1405810@qu.edu.qa;f.bensaali@qu.edu.qa} \\
   \And
 Abbes Amira \\
  Institute of Artificial Intelligence\\
  De Montfort University\\
  Leicester, United Kingdom \\
  \texttt{abbes.amira@dmu.ac.uk} \\
}

\begin{document}
\maketitle

\begin{abstract}
Recently, tremendous interest has been devoted to develop data fusion strategies for energy efficiency in buildings, where various kinds of information can be processed. However, applying the appropriate data fusion strategy to design an efficient energy efficiency system is not straightforward; it requires a priori knowledge of existing fusion strategies, their applications and their properties. To this regard, seeking to provide the energy research community with a better understanding of data fusion strategies in building energy saving systems, their principles, advantages, and potential applications, this paper proposes an extensive survey of existing data fusion mechanisms deployed to reduce excessive consumption and promote sustainability. 
We investigate their conceptualizations, advantages, challenges and drawbacks, as well as performing a taxonomy of existing data fusion strategies and other contributing factors. Following, a comprehensive comparison of the state-of-the-art data fusion based energy efficiency frameworks is conducted using various parameters, including data fusion level, data fusion techniques, behavioral change influencer, behavioral change incentive, recorded data, platform architecture, IoT technology and application scenario. Moreover, a novel method for electrical appliance identification is proposed based on the fusion of 2D local texture descriptors, where 1D power signals are transformed into 2D space and treated as images. The empirical evaluation, conducted on three real datasets, shows promising performance, in which up to 99.68\% accuracy and 99.52\% F1 score have been attained. In addition, various open research challenges and future orientations to improve data fusion based energy efficiency ecosystems are explored. 
\end{abstract}


\section{Introduction} \label{sec1}
Recently, many governments have put energy efficiency among their highest priorities since energy preserving is the simpler way to save money for consumers, reduce CO$_{2}$ emissions and further reduce dependency on fossil fuels. To this end, a set of binding targets to reduce energy consumption has been compiled aiming at improving the energy saving rate by 2030 by at least 32.5\%, relative to a \enquote{business as usual} scenario \cite{HEUN2019112697,Bisegna8783774}. As households and public buildings are reported to consume more than 40\% of energy and can generate more than 35\% of CO$_{2}$ emissions, reducing their energy consumption and enhancing their performance has become a principal goal aiming at meeting long-term energy efficiency, climate and environment improvement goals \cite{DAGOSTINO2019100412}.

To that end, a high priority has been given to set powerful energy efficiency strategies, where significant investments have been made to develop new technologies and materials to improve energy efficiency. However, there are various human-related elements and actions that must be considered and corrected in order to improve energy saving in buildings \cite{Mahapatra2018,MAMOUNAKIS2019130}. Put simply, additional efforts need to be devoted to increase energy awareness and hence promote behavioral change among consumers to reduce wasted energy \cite{Sardinos2019,Alsalemi2018IEESyst}. Moreover, it is also of urgent need that strategies and targets should be in accordance with the preferences of buildings' end-users and owners, and further actions should be smoothly embedded into daily behaviors in order to be effective \cite{DELZENDEH20171061,ORTIZ2017323}. It is worth noting that end-users' behavior is responsible of wasting more than 20\% of the total energy consumed in buildings, and hence it is a key element in energy consumption \cite{White2019,URGEVORSATZ201585,ALMARRI20173464}. Consequently, it is of paramount importance to develop novel platforms and tools to gather power consumption data, which will help end-users understand their energy consumption footprints and improve their consumption behavior \cite{Paone11040953}.

In this line, daily behavioral change is a great challenge requiring that end-users should be trained and awarded for their efficient consumption actions through providing them with continuous feedback data and incentives. This results in promoting and motivating them to achieve a long-term behavioral change. To this end, there is a rising interest to the behavioral change topic, and especially to its application in energy efficiency \cite{Varlamis2020CCIS,ZHAO2019113701}. Consequently, in this paper, more importance is paid to reviewing data fusion based energy efficiency systems that are mainly from the consumer's perspective, i.e. related to behavioral change. In addition, it is worth mentioning that improving end-users consumption behavior has lately emerged as a cost-effective strategy to reduce wasted energy in buildings \cite{DELZENDEH20171061,STADDON201630,RAFSANJANI2018317,ORNAGHI2018582}.

Energy efficiency based on behavioral change relies mainly on analyzing power consumption footprints and other contextual information in order to provide end-users with tailored recommendations on how to optimize their energy usage \cite{YIN2020119234}. End-users will then follow these recommendations manually or using Internet of things (IoT) control techniques to achieve greater energy savings \cite{SPANDAGOS2020115117,SAFARZADEH201944}. Aiming at formulating personalized recommendations for end-users, data collected from different sensors and multiple differing modalities are aggregated, analyzed and classified in order to detect abnormal energy usage before generating appropriate decisions \cite{ZHANG202065,KHALEGHI201328,DING202084}. In this regard, developing schemes to acquire data, learn and manage associated knowledge is a must to support the interconnection between gathered footprints from different sources and modalities with the available information sources (historical data), and design appropriate information fusion strategies in accordance with the underlined goals \cite{DENG201990}. It is also important to take into consideration the complex event processing \cite{DEFARIAS2019109} solutions to correlate data coming from different data streams in different formats. To that end, data fusion strategies that play an essential role in producing reliable and accurate information have been extensively investigated \cite{Fotopoulou8016276}.

In the literature, several solutions have been developed aiming at exploring  the potential introduced by various novel information and communication technologies (ICT) to design effective energy efficiency systems based on consumers' behavioral change. These ecosystems have been developed using IoT technologies, sub-meters, smart sensors, aggregation and fusion techniques. Moreover, other procedures have been also deployed, including mobile recommender systems, serious gaming \cite{Daniel2017,Casals2017a} and machine learning (ML) models that promote end-users' behavioral change. This can be achieved by (i) providing them with personalized recommendations to reduce wasted energy (ii) motivating, engaging and educating them regarding energy consumption and related concerns, and (iii) measuring their awareness, attitudes, engagement and self-reported energy efficiency behaviors.

Further, the energy efficiency systems have widely been affected by the rapid development of the IoT technology, which is expected to connect all items (such as, households appliances, mobile devices, sensors, etc.) enabling them to exchange data. IoT has an important role in different energy efficiency ecosystems for remote acquisition and control. In order that energy efficiency systems benefit from IoT technology; collecting and recording data of indoor and outdoor ambient conditions, energy consumption, end-user's habits and preferences are the first steps towards a powerful energy saving framework. Different kinds of sensors can be used with the view of improving the breadth of gathered data and enhancing the depth of information fusion results since a unique sensing modality is not enough to design an effective energy efficiency ecosystem \cite{LIU2020123,PENG2020199}.

Consequently, a multi-perspective study to investigate the relevance of information fusion in energy efficiency systems is required through broadening scale and scope of data sources, information record schemes, data pre-processing approaches and fusion strategies. Specifically, for developing smart automatic energy efficiency frameworks, data from different sensor modalities are utilized, where their performance should be evaluated based on various parameters. In this regard, a multi-aspect study that aims at covering the entire depth and breadth of information fusion issues in energy efficiency ecosystems is proposed. These aspects are related to information fusion goals, information fusion methods, modalities of data sources, nature of output data, information fusion levels, and platform architectures for behavioral change. 
In this investigation, we provide a taxonomy of data fusion strategies used in energy efficiency frameworks at different levels, including multi-sensor data fusion, fusion of appliance features and fusion of semantic data. For every category, we only review notable works with the object of demonstrating the generality and effectiveness of our multi-perspective scheme to evaluate data fusion approaches. We provide also a generic classification of energy efficiency frameworks based on various perspectives including (1) data fusion level; (2) data fusion techniques; (3) behavioral change influencer; (4) behavioral change incentive; (5) recorded data; (6) platform architecture; (7) IoT technology; and (8) application scenario. 

Furthermore, a comparison between several energy efficiency ecosystems is carried out as well based on the aforementioned aspects. In addition, an example of applying data fusion to an appliance identification purpose is presented, where a novel scheme based on the fusion of two-dimensional (2D) descriptors is used. This method relies on deploying 2D descriptors to extract relevant appliance fingerprints, then a fusion is introduced in order to improve the accuracy of appliance identification. To the best of the authors knowledge, this is the first work that discusses the applicability of 2D descriptors to extract features from one-dimensional (1D) power signals of electrical devices. 

Afterward, we discuss open research issues and valuable future orientations of data fusion based energy efficiency frameworks, which are useful for the energy efficiency research community. To summarize, this paper presents a set of novel contributions, which can be listed as follows:
 
\begin{itemize}
\item Overviewing valuable state-of-the-art data fusion based energy efficiency works. 
\item Proposing a multi-perspective taxonomy to assess existing information fusion schemes and energy efficiency ecosystems.
\item Presenting a survey of energy efficiency ecosystems with a comprehensive comparison among them and describing the current trend of information fusion strategies in each framework.
\item Proposing a novel appliance identification technique based on the fusion of 2D descriptors.
\item Providing a list of open research issues and future orientations for improving information fusion based energy efficiency systems.

\end{itemize}

The remaining parts of this paper are structured as follows. In Section \ref{sec2}, we present an overview of state-of-the-art data fusion and energy efficiency related-works and their taxonomy based on several development stages. Moreover, a comparison between a set of existing energy efficiency ecosystems is conducted based on the aforementioned features. Following, Section \ref{sec3} presents an example of data fusion strategies use for appliance identification, in which a novel scheme is proposed based on the fusion of 2D descriptors to extract pertinent features from power consumption signals of electrical devices. In addition, the performance of the proposed method is evaluated on three open access datasets with reference to various ML classifiers. In Section \ref{sec4}, a set of open research issues are presented with valuable future orientations and perspectives are drawn. Conclusions and recommendations are provided in Section \ref{sec5}, while acronym definitions of the energy efficiency frameworks discussed in Section \ref{sec2} are lastly described in the appendix.

\section{Overview of data fusion and energy efficiency} \label{sec2}
\begin{figure}[t!]
\begin{center}
\includegraphics[width=17.6cm, height=7.9cm]{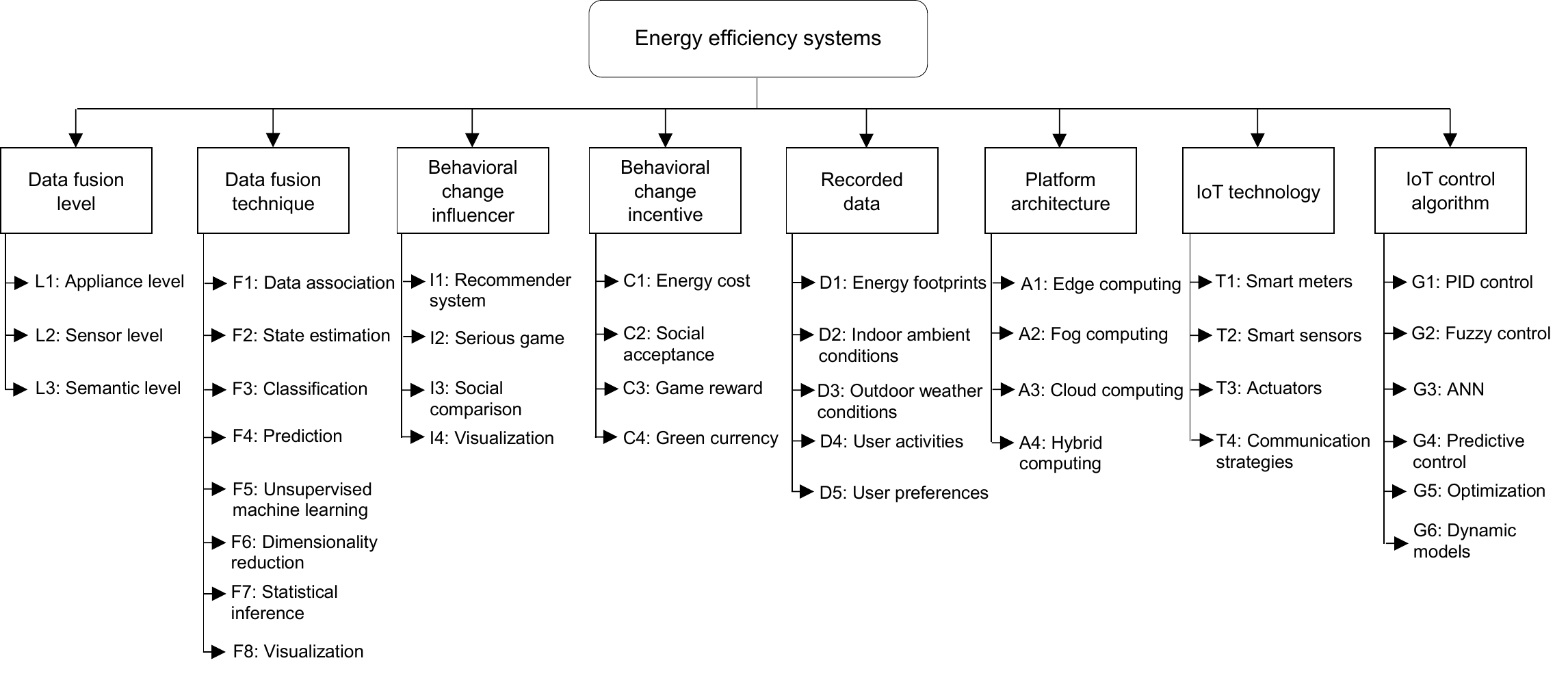}
\end{center}
\caption{Taxonomy of energy efficiency ecosystems based on data fusion and other parameters.}
\label{EEtaxonomy}
\end{figure}

In order to conduct a comprehensive review, a taxonomy of data fusion based energy efficiency related-works is performed, in which we describe various generic perspectives aiming at covering the entire depth and breadth of state-of-the-art information fusion based energy efficiency systems. 
The proposed taxonomy encompasses eight principal categories: data fusion level (L), data fusion technique (F), behavioral change influencer (I), behavioral change incentive (C), recorded data (D), platform architecture (P), IoT technologies (T) and IoT control algorithms (G). These categories are adopted in this framework since we review the energy efficiency systems from the consumer's perspective. Therefore, more importance has been given to the strategies that help in improving the consumer's behavior, such as behavioral change influencer and behavioral change incentive. Moreover, since data fusion can be  applied in different levels and using various techniques, it is of utmost importance to apply a taxonomy in this direction. In addition, the data collection step is of great significance for every energy efficiency system since it allows data collection in real-time and with better quality. Consequently, reviewing recorded data, platform architectures and IoT technologies is a must. Finally, a brief description of the IoT control algorithms is also provided since they help the consumer to automatically monitor electrical devices, promote behavioral change and hence reduce wasted energy. Fig. \ref{EEtaxonomy} indicates a complete list of the adopted taxonomy illustrating all the classes and their sub-classes. 

\subsection{Data fusion level (L)}

In energy efficiency literature, there are three main application levels of data fusion, which are multi-sensor fusion, appliance features fusion and semantic data fusion. 

\begin{itemize}
\item \textbf{L1. Appliance level fusion}
\end{itemize}

One important application of data fusion strategies is related to appliance identification and non-intrusive load monitoring (NILM). This application provides promising insights since the power consumption of each electrical device can be inferred from the aggregated energy consumption record without the need to install a sensor for each appliance \cite{HIMEUR2020114877}. Therefore, this considerably reduces the implementation cost of energy saving ecosystems. In addition, different feature fusion techniques can be deployed to extract appliance fingerprints, and thereby help in improving the appliance identification accuracy. 

In \cite{Himeur2020}, a non-intrusive appliance recognition method providing particular consumption footprints of each appliance is proposed. Electrical devices are recognized by the combination of different descriptors via the following steps: (a) investigating the applicability along with performance comparability of several time-domain (TD) feature extraction schemes; (b) exploring their complementary features; and (c) making use of a new design of the ensemble bagging tree (EBT) classifier. Consequently, a powerful feature extraction technique based on the fusion of TD features (FTDF) is proposed, aimed at improving the feature discrimination ability and optimizing the recognition task.

In \cite{Phillips2013}, a multi-sensor fusion technique is implemented to infer appliance-level power consumption. In this regard, appliance features extracted from various domestic devices after the event detection stage are fused before applying various unsupervised ML algorithms. The multi-sensor fusion aims mainly at correlating data gathered by energy consumption submeters, light, and acoustic sensors in order to reduce possible sensing errors. 

In \cite{Pathak2015}, an NILM method is developed using the fusion of acoustic and power signatures. To that end, Pathak et al. correlate an electrical device's inherent acoustic noise with its power usage footprints separately and in presence of other devices. Following, various classifiers are investigated to model the relation between device energy consumption and acoustic
states. This aids in providing an efficient appliance identification and detecting abnormal events. \vskip2mm

\begin{itemize}
\item \textbf{L2. Sensor level fusion}
\end{itemize}

Occupancy detection in buildings supports reducing wasted energy and personalizing comfort management. In fact, occupancy-driven management helps saving more than 15\% of the total energy consumption in buildings \cite{Yuvraj2010}. End-users can then be provided with appropriate energy saving recommendations based on the analysis of occupancy patterns. However, from analyzing existing occupancy detection frameworks, it is clearly seen that in order to collect real-time occupancy data in a household, data fusion of various implemented, not-necessarily-smart, yet accurate enough sensors is required. The sensor fusion aims at making the task of occupancy detection smart without the need of user intervention.

In \cite{Ekwevugbe2012}, a multi-sensor fusion technique is described to detect occupancy in household building using an adaptive neuro-fuzzy inference system (ANFIS). This framework records ambient conditions, indoor  events  and  power consumption patterns collected from a domestic household to gather occupancy footprints. In \cite{Ekwevugbe2013}, a multi-sensor fusion technique is proposed, which combines multiple types of data, including sound degree, indoor temperature, CO$_{2}$ level and motion patterns for estimating occupancy footprints. An infrared camera is also deployed to detect occupancy rates. To extract features, the authors deploy symmetrical uncertainty test and a genetic-based method for evaluating an optimum sensor aggregation. Further, the information fusion procedure is conducted using a neural network.

In \cite{Nasir2015}, an occupancy detection system in an office space is proposed. In this regards, a smart-door prototype has been designed through the combination of data analytic patterns and multi-sensor fusion. This helps monitoring building with a high occupancy level through managing the indoor conditions and hence reducing energy consumption, especially due to air-conditioning and heating systems. In \cite{Hsiao2015}, the status of house occupancy is inferred using a multi-sensor fusion. Specifically, spatial and temporal patterns are combined based on a fusion model. This technique resolves the issue of losing sensing values of pyroelectric infrared (PIR) sensors, and hence enhancing the occupancy detection. 

In \cite{Woodstock7528087}, Woodstock et al. propose a multi-sensor fusion testbed incorporating a sparse matrix of time-of-flight range sensors and ML models for occupancy detection. Accordingly, a multi-sensor fusion is proposed using the combination of data from the matrix of 18 time-of-flight sensors and associated techniques linking sensor footprints to occupancy and activity detection. In \cite{CHEN2016790}, different ML models are used to build a household occupancy detection using environmental sensor patterns. The outcomes of those ML schemes are then fused to improve the occupancy detection accuracy. Afterwards, occupancy-patterns are used to automatically control the energy usage and thereby help saving up to 15\% of wasted energy.

In \cite{Sangoboye2016}, an occupancy-driven control is implemented to accurately monitor occupancy data through the fusion of multi-sensor count lines. To that end, a probabilistic model is applied using occupancy constraints to accurately estimate the number of occupants.  
In \cite{Fiebig2017}, an occupancy detection framework is implemented to improve the energy management in a domestic building. Footprints collected from six air quality sensors are used along with a set of semantic data that represent calendar entries to estimate the number of occupants in a pilot building. Recorded data are then preprocessed before extracting the features that are then fed into a set of multi-classifier models, including multilayer perceptrons (MLP), k-nearest neighbors (KNN), decision trees (DT) and random forests (RF). Following, a fusion strategy is applied on the results obtained from those classifiers to detect occupancy level.

In \cite{K-DAS2017}, occupancy patterns are identified in a university campus using data fusion of three sensor modalities, including energy consumption, characteristics of (wireless fidelity) WiFi sensors and rate of water usage. After extracting the features from these modalities, a data fusion stage is applied before deploying a clustering mechanism for identifying similarity in buildings' features and training per cluster occupancy estimation models. 
In \cite{Pratama2018}, a low-intrusive occupancy detection system in a typical workspace is proposed, which is built upon the fusion of data collected from each office energy consumption at aggregated and appliance levels and applications running on workers' mobile devices. The received signal strength (RSS) of Bluetooth low energy (BLE) nodes utilized in the workspace are used to pinpoint phone devices in each office.
In \cite{WANG2018233,WANG2019106280}, to design an optimal occupancy detection model, Wang et al. combine three modalities of data, which represent ambient footprints, WiFi data and fused environmental sensing data. Three ML models are also investigated, including kNN, support vector machine (SVM) and artificial neural networks (ANNs). 
In \cite{FAYED2019e02450}, to detect whether a room is occupied or not, Fayed et al. propose a heterogeneous data fusion from different sensor modalities. Specifically, they use neutrosophic ensembles with the correlation of multi-sensor fingerprints.

In \cite{Nesa2017}, a remote monitoring of energy consumption in a building through heating, ventilation, and air conditioning (HVAC) control is achieved based on detecting occupants. Specifically, occupancy detection is realized using the fusion of ambient conditions, including temperature, humidity, light and CO$_{2}$. While in \cite{HOBSON2019106154}, HVAC monitoring is managed using an occupancy-count estimation via the fusion of multi-sensor patterns. This framework considers data from different sensor modalities such as WiFi parameters, CO$_{2}$ information, PIR motion detection values, plug and light power consumption footprints.
In \cite{WANG2019386}, an information fusion framework is proposed to predict energy usage gains achieved through the heating system using a deep learning model. In practice, long short-term memory (LSTM) networks are used to predict heating gains using multiple categories of data from various sensors, including miscellaneous electric loads (MELs), lighting, occupancy patterns, WiFi connection counts.
In \cite{GUARINO2019240}, data collected from different sensor modalities are fused to develop a smart energy usage system. The latter can manage energy consumption of cooling and heating systems in a reference building with respect to expected climate changes. Two fusion strategies are deployed using the correlation between errors of punctual historical patterns or prediction models and real data. Consequently, obtained features are then utilized to predict heating and cooling power requirements. In \cite{LI2016227}, a data fusion strategy based on Kalman filter is proposed to predict energy consumption in a typical building. The developed model is then used to monitor various electrical appliances in a realistic scenario. The fusion is performed to achieve an enhanced energy efficiency behavior using data from multiple modalities, including outdoor air temperature, lighting patterns, indoor temperature and humidity, direct solar radiation and ventilation rate. \vskip2mm

\begin{itemize}
\item \textbf{L3. Semantic level fusion}
\end{itemize}

Also called decision level fusion, it mainly refers to the fusion of high-level information, which are usually linked to decision generation. The data fusion step is performed to help triggering final decisions based on aggregating/combining multiple sub-decisions. In comparison to the low-level fusion, semantic fusion schemes generally produce an initial classification based on the use of various information types and hence can achieve high fusion accuracy.

In \cite{Wun2007}, Wun at al. propose a high-level semantic data fusion scheme occurring through different sensor networks along with aggregating data within every single network.
In \cite{Zafeiropoulos2008}, a semantic data fusion architecture is introduced helping in aggregating, enriching, managing and querying data from various sensor modalities. A smart energy efficiency solution is then designed through extracting meaningful patterns from the sensor nodes raw data.
In \cite{Fensel2012}, Fensel at al. propose the SESAME framework that is based on the fusion of ontology-based data, aiming at reducing wasted energy and describing the relation between IoT nodes and consumers within domestic buildings. 
In \cite{Noughabi2013}, in order to surmount semantic issues in energy efficiency based heterogeneous systems, a semantic data fusion technique is developed using the Joint Directors of Laboratories (JDL) fusion strategy. 

In \cite{Terroso2016}, two semantic models are developed and fused in order to portray electricity usage and behavioral properties of end-users and further reduce energy consumption via influencing their consumption behavior. This has been achieved using personalized recommendations provided through a specific gamification application.
In \cite{BouryBrisset2003},  the contribution of ontology based approaches in helping data fusion from multiple information modalities is discussed. Specifically, a systematic scheme supporting high-level data fusion is proposed to promote a better semantic information integration and hence developing a powerful energy efficiency ecosystem.
In \cite{Xiao}, an ontology method to model information fusion, exchange and further semantic data retrieval is proposed for a better energy management in buildings. In this regard, semantic information collected from multiple data sources and ontologies are then deployed for representing energy consumption states and describing functionalities of a developed energy efficiency system.
In \cite{Fotopoulou2017c}, an energy efficiency ecosystem is implemented using an information fusion mechanism. It comprises four main steps defined as (a) collecting data from different sensor modalities, (b) representing them in a semantic space, (c) defining semantic models, and (d) applying various data fusion strategies to come with efficient and personalized energy saving measures. 
In \cite{Jing2020}, high-level decisions are generated in order to reduce wasted energy on campus based on a multidimensional situational data fusion technique. The latter aims at normalizing, analyzing and predicting different information over multiple source modalities, including natural, humanistic and spatiotemporal footprints.

\subsection{Data fusion techniques (F)}
In this section, we overview common information fusion schemes. From F1 to F3, conventional data fusion approaches are discussed. From F4 to F8, data fusion associated with ML are outlined, in which simple input footprints from different sensor modalities are fused to result in a higher level data enrichment. In what follows, we summarize the characteristics of these categories:

\begin{itemize}

\item \textbf{F1. Data association:} represents data fusion schemes fusing data using the correlation between at least two or more information sources. Conventional schemes for data association incorporate KNN \cite{FAYED2019e02450}, probabilistic data association (PDA) \cite{Furqan2017,AbuBakr2017}, and multiple hypothesis test (MHT) \cite{Wijayasekara2015,Izumi2018}.

\item \textbf{F2. State estimation:} refers to the use of information from multiple sensors in different modalities to reach a high state estimation accuracy. Practical methods pertaining to this class include maximum likelihood estimation (MLE) \cite{Manish2005}, Kalman filter \cite{LI2016227,Xiwang2014}, particle filter \cite{HU2015229}, and covariance consistency model \cite{Jeffrey2003,Wangyan2015}.


\item \textbf{F3. Classification:} indicates the strategy of clustering data into distinct categories using their unique characteristics. More discussion about generic classification approaches can be found in \cite{Fiebig2017,Wijayasekara2015}.

\item \textbf{F4. Prediction:} methods belonging to this class are mainly utilized to predict outputs based on single or multiple information sources. It is worth noting that this class deals with simple techniques, e.g. regression and complicated approaches, e.g. prediction modeling. An example of some frameworks using data fusion based on prediction can be found in \cite{WANG2018233,WANG2019106280}

\item \textbf{F5. Unsupervised machine learning:} this category refers to the kind of fusion schemes that aims at automating knowledge discovery without the need of having the ground-truth labels. To this end, clustering \cite{Oyaga2017}, anomaly detection \cite{Xingfeng2015} and other strategies \cite{Lin2009ieee} are widely used. It is noteworthy to mention that semi-supervised ML algorithms \cite{Larios2012} are categorized under this group as well.

\item \textbf{F6. Dimensionality reduction:} refers to algorithms attempting to decrease the dimensionality of data gathered from various sensor modalities while decreasing the computation time  to process high-dimensional data. They are used to develop powerful feature extraction modules or visualization tools. Principal component analysis (PCA) \cite{AZADEH20073792,Zhang2012c,Hogun2014} is among the well-known dimensionality reduction schemes used in energy efficiency frameworks in addition to others \cite{Han2011}. 

\item \textbf{F7. Statistical inference:} schemes pertaining to this class are utilized to outline some specific features in addition to some common knowledge/hypothesis extracted from energy monitoring sensors. The NILM (named also energy disaggregation) and appliance identification tasks are among the principal applications of such schemes. A large number of papers based on these techniques can be found in the literature such as \cite{BRAULIOGONZALO2016198,PARSON20141,BASU2015109,Makonin7317784,Guedes2015,KRULL2018119,Kong7750555,Liu7795164,Ji8684887,Mengistu8337762,HOLWEGER2019100244,BONFIGLI20171590,COMINOLA2017331,Henao11010088,Raiker8721436,Song8483075,Kong7508980}.

\item \textbf{F8. Visualization:} includes techniques used for presenting energy consumption statistics of buildings and their appliances to consumers through the use of appropriate platforms. To design a powerful energy consumption visualization tool, multiple data sources are fused in order to provide  end-users with relevant information about the factors that affect their energy usage. A set of visualization tools can be found in the following frameworks \cite{FRANCISCO2018220,SMITH201968,TORABIMOGHADAM201919,ABDELALIM2017258}.

\end{itemize}

\subsection{Behavioral change influencer (I)}
Energy consumption behavioral change is among the main strategies used to influence and improve the energy usage behavior of end-users in order to reduce wasted energy and optimize energy management. Various techniques have been proposed to achieve and support this purpose. These methods usually merge the empirical interpretation of consumers' behavior with the quantization of power efficiency measures aiming at fostering energy saving.

\begin{itemize}
\item \textbf{I1. Recommender system (RS):} a recommender system is an engine that is used to help users planning their electricity consumption more easily, through providing them with pertinent and actual data recommendations. It is based on investigating ways to evaluate what kind of data are appropriate and useful to make decisions about power consumption \cite{Schweizer2015,Alsalemi2018IEESyst}. Then, personalized recommendations are generated and transmitted to end-users via a set of notifications. Different sorts of recommender systems are proposed in the literature using the input data either to simply select actions of possible interests to the target consumer or to predict the consumer interest level for specific actions and then produce appropriate recommendations. Consequently, several recommendation engines are proposed such as collaborative filtering \cite{Zhang8412100,Sardinos2019}, context-aware recommendations \cite{Weng6385762}, content-based recommendations \cite{Chen2016,Alsalemi2019ISA} and multi-agent recommendations \cite{ALDARAISEH2015958,Diego2019}.

\item \textbf{I2. Serious game (SG):} the deployment of serious games, also called of gamification to boost power saving behavior of building occupants is a hot research topic \cite{Casals2017a,CASALS2020109753}. It is used to improve the learning and involvement of end-users to decrease domestic wasted energy \cite{ALSKAIF2018187,Casals2016,JOHNSON2017249}.

\item \textbf{I3. Social comparison:} various energy efficiency frameworks have been implemented based on incorporating a normalized comparison (consumer ranking) module via an eco-feedback tool allowing consumers to compare their power consumption footprints with their peers and neighbors \cite{MORLEY2018128,WEMYSS201916,DU201656}. This influncer engine is successful because it depends on the assumption that consumers are highly marked by engagements and rankings of others in their social networks \cite{JAIN2013119}. The works conducted in this direction have revealed the effect of interactions between consumers via social networks for a significant reduction of wasted energy \cite{PESCHIERA2012584}.

\item \textbf{I4. Visualization}: how to inform consumers using mobile phones and IoT devices to obtain real-time statistics of their energy consumption footprints has a major influence on reducing wasted energy in buildings and improving consumers' habits of electricity usage \cite{Pahl2016,Spence2018,Rist2019,Herrmann2018}. To that end, it has been demonstrated in multiple papers\cite{Alsalemi2019ISA,ABDELALIM2015334,MURUGESAN20151,ELBELTAGI2017127,ELLEGARD20111920} that inventive visualizations of real-time electricity usage fingerprints prompt more sustainable behavior. A creative and interactive visualization can display the current energy consumption of all electrical devices offering new measures for conserving energy at homes and workplaces \cite{Holmes2007,Costanza2012}.

\end{itemize}

\subsection{Behavioral change incentive (C)}
Behavioral change incentive refers to the tools that are used to motivate end-users behavioral change and increase their awareness. They are implemented to permit the exchange of benefits, feedback and experience generated to award the end-users' efficient behavioral change.

\begin{itemize}

\item \textbf{C1. Energy cost:} in this case, triggering users to accept recommended energy-saving measures is done through providing them with financial cost  reduction based on their actions \cite{CASALS2020109753}.  

\item \textbf{C2. Social acceptance}: in this case, ratings related to users' awareness and social acceptance of the energy saving measures established by a specific energy efficiency solution are measured. The feedback is then transmitted to these users in order to motivate them in developing more sustainable actions \cite{Dawid2017}.

\item \textbf{C3. Game reward:} the game reward is mainly linked to the case in which the influencer system is built upon a serious game. In this context, an ensemble of services are established to measure game rewards related to real-world power saving orientations \cite{CASALS2020109753}.

\item \textbf{C4. Green currency:} to endorse energy saving of householders and building occupants using the consumption feedback, some energy efficiency frameworks introduce new currencies such as the CO$_{2}$ currency as a stimulus to promote sustainability \cite{JOACHAIN201489}.

\end{itemize}

\subsection{Recorded data (D)}
To implement an energy efficiency system, various kinds of data are recorded from different sources. Generally, there are five generic types of recorded data that can be classified without referring to the sensor or communication medium deployed to collect them. They can be summarized as follows:

\begin{itemize}
\item \textbf{D1. Energy consumption footprints:} they represent energy consumption fingerprints collected form specific building either at the aggregated-level or appliance-level \cite{Cuimin2019,Shin2019,Himeur2020icict}.

\item \textbf{D2. Indoor ambient conditions:} this data group deals with the collection of indoor ambient patterns, including temperature, humidity, luminosity, CO$_{2}$ emissions collected inside buildings \cite{Nesa2017}.

\item \textbf{D3. Outdoor weather conditions:} in this class, outdoor environmental patterns are gathered from online sources to incorporate more contextual data into the system. This kind of data includes temperature, humidity, wind speed/direction and solar radiation \cite{Moon2011,Fikru2015}.

\item \textbf{D4. User activities:} this data category refers to information related to user activities. Specifically, sensors are placed to monitor user's physical activity such as occupancy \cite{WANG2018233,WANG2019106280}, thermostat adjustments \cite{Moon2011}, windows and doors openings \cite{Heeboll2018}.

\item \textbf{D5. User preferences:} in order to develop a powerful energy-efficiency solution, user preferences, comfort-related practices, and habits are among the important information that should be taken into consideration \cite{Paone2018, MOBISTYLE2019}.

\end{itemize}

\subsection{Platform architectures (A)}
The platform architecture of data fusion based energy efficiency systems represents another essential classification context. In this class, four general groups are identified:

\begin{itemize}
\item \textbf{A1. Edge computing architecture:} this architecture platform refers to the case in which information processing and fusion are conducted at the edge level. This is done in the physical layer, where multiple kind of information are actually gathered. This architecture makes use of different platforms, including micro-controllers, computing platforms (Arduino, Raspberry Pi) \cite{RPI4}and multi-core embedded platforms (such as Jetson Nano \cite{JetsonNano}, Jetson TX1 \cite{JetsonTX1}, Jetson TX2 \cite{JetsonTX2} and ODROID \cite{ODROID}). More discussion on the implementation of such architecture in energy efficiency ecosystems can be found in the following papers \cite{Candanedo2019,MOCNEJ2018162}.

\item \textbf{A2. Fog computing architecture:}  in this architecture, the information processing and fusion are conducted in the middle layer situated between the edge and the cloud. It relies on a periodic and continuous information sampling at the edge level without processing before forwarding the sampled data to a gateway, which is acting as a fog device \cite{Toor2019b,Tian2016}. Following, information processing is performed at the gateway level where computing materials are available. Usually, fog and edge computing architectures supply similar advantages of offloading computing as discussed in \cite{OMA201814,Toor2019}. 

\item \textbf{A3. Cloud computing architecture:} in this case, information fusion and processing is managed at the cloud level. Using the cloud platform is considered as the most practical way to process big data, either in industrial or academic sectors \cite{Zirnhelt2014}. Cloud computing presents a set of benefits such as the immediate access to information sources on both online and offline manners in order to conduct further processing and fusion. The main limitations of the cloud computing are related to the increased cost and communication overheads. Considerable interest has been paid to the use of cloud platforms to develop energy saving frameworks \cite{CHOU2016397,Ferrandez2019}. 

\item \textbf{A4. Hybrid computing architecture:} it concerns the case in which data processing is divided into different layers such as the edge, fog and cloud as mentioned in \cite{Zhang2017}. For this case, based on the available computation material and application aims, low-level information fusion and/or processing can be performed either at the edge and/or fog, whereas high-level data management can be realized at the cloud \cite{Anjomshoaa2018,Izumi2018}.

\end{itemize}

It is worth noting that the cost of the energy saving solution can be configured based on the selected platform architecture, which is mostly related to
the specific application scenario. For example, edge computing architectures allow developing low-cost energy efficiency solutions.

\subsection{IoT technologies (T)}
IoT is a technology allowing objects and devices connected via a network to communicate between them, including smart meters, smart sensors, actuators and processors. This helps in providing different kinds of services and further contributing in saving time, reducing cost and allowing a real-time remote data collection \cite{KHAJENASIRI2017770,POCERO201754}. In IoT-based energy efficiency systems, different sorts of smart sensors are deployed to sense data from different modalities. Collected data are then transmitted to control centers with the aim of storing, processing, analyzing and generating recommendations or making decisions to reduce wasted energy. In case of automatic energy management, tailored commands are sent back to actuators installed in buildings as a result of the analysis made using sensed data. Because diverse kind of sensors, actuators and communication techniques are proposed in the literature, we describe briefly exiting technologies enabling IoT-based energy efficiency in buildings \cite{MARTINLOPO2020107101}.

\begin{itemize}
\item \textbf{T1. Smart meters:} they allow remote power consumption collection of aggregated and/or disaggregated (appliance level) circuits in different buildings \cite{Dharur8058288}. In addition, they allow a real-time monitoring of consumption patterns, and hence help in accelerating the analysis process \cite{MOGLES2017439,BHATI2017230}. This results in identifying problems and anomalies for actual consumption data.   
\item \textbf{T2. Smart sensors:} are the key elements in IoT energy efficiency systems. They are employed to collect and transmit various kinds of data such as ambient conditions, occupancy and luminosity \cite{ALSALEMI8959214,Hu8423630}. Specifically, the deployment of smart sensors helps in enhancing the effectiveness, functionality and interoperability of IoT-based energy efficiency systems \cite{Marinakis2018,TANG2019127}. Several types of smart sensors exist for different purposes such as: 
\begin{itemize}
\item Temperature sensors: are deployed to sense indoor temperature levels, which can be used to adjust the energy usage of heating and cooling equipment inside buildings \cite{Akmandor8432098,Alsalemi2020ieeeSyst}.
\item Humidity sensors: they are deployed to detect the amount of moisture and air's humidity inside buildings. This kind of information is important to adjust the energy consumption of cooling and heating systems as well \cite{Tushar2016}.
\item Luminosity sensors: are deployed to measure luminosity (ambient light level) inside buildings. The feedback of these sensors is then used to control the power consumption of light lamps in buildings \cite{s17102296}. They are very important in energy efficiency systems because lighting accounts for more than 15\% of the overall energy consumption in buildings \cite{en13020494}.
\item Occupancy sensors: are used to detect the presence of end-users inside the buildings. This information is relevant since some appliances should turn off if the consumers are absent, this is the case for example of air conditioners, televisions, fans and heating systems. Therefore, this information is very helpful to detect abnormal consumption related to the presence/absence of end-users \cite{Akkaya7122529,GARG200081,HimeurCOGN2020}. 
\end{itemize} 
\item \textbf{T3. Actuators:} are the tools used for transforming certain forms of energy into physical manifestations. They receive electrical inputs from energy efficiency systems, transform them into specific actions, and hence acting on the electrical appliances connected to the IoT network to reduce their energy consumption, e.g. turning off an appliance \cite{Zhao8885095}. Different kinds of actuators have been used in IoT-based energy saving systems, such as electric actuators \cite{Sarwesh7380569}, hydraulic actuators \cite{en13020494}, pneumatic actuators \cite{Luo6236193} and thermal actuators \cite{en13020494}.   
\item \textbf{T4. Communication strategies:} 
they have a significant role in connecting electrical appliances in buildings into the IoT network. Wireless communications are widely used to connect appliances to IoT gateways and carry out end-to-end data communication between the different elements of the energy efficiency system \cite{Metallidou2020}. Wireless systems are implemented  with regard to various wireless protocols and the utilization of every element relies on the requirement of the energy efficiency system, including communication range, bandwidth, and power usage specifications \cite{Moreno7248403,Hannan8403212}. To transmit the collected power consumption footprints and ambient data, short range wireless technologies have been first adopted, such as WiFi. However, for the energy efficiency application, Wi-Fi is not the best option because of their power requirements \cite{Arshad7997698}. Consequently, other low-power wide area network (LPWAN) solutions have successfully replaced Wi-Fi such as:
\begin{itemize}
\item Bluetooth Low Energy (BLE): is a short-range wireless communication solution enabling transmission of collected data based on short radio wavelengths. Moreover, it a cost effective solution that has an average range of 0--30 m. Various energy efficiency works have adopted this technology to transmit collected data, such as \cite{Choi7066499,en81011916,WANG20181230}.
\item Zigbee: is developed to implement personal area networks and transmit collected data in small buildings. Zigbee is a cost-effective solution, which is simple to implement and can ensure a low-data rate with a highly reliable network, especially for implementing building energy efficiency systems with low-power requirements \cite{PARK2013662,SINGARAVELAN2016305}.
\item Long Range (LoRa): is another low-cost communication solution deployed for IoT-based energy saving systems in large buildings. It is mainly adopted to transmit data for long-distances, e.g. up to 10 km and can ensure a battery life for several years, mainly due to its very low power consumption \cite{Kam2019,electronics8091040}.   
\item Narrowband IoT (NB-IoT): is another LPWAN communication solution, which connects a large number of appliances, smart meters and smart sensors and allows high data transmission rates along with low latency. Moreover, NB-IoT is a also a cost-effective strategy with long battery life, which has been widely deployed in energy efficiency systems \cite{Pirayesh8737625,Popli2018,SONG2017460}. 
\end{itemize}  
\end{itemize}

\subsection{IoT control algorithms (G)}
To apply the recommended actions generated by the recommender systems or learned using serious games in an automatic manner, it is required to use efficient IoT control algorithms to monitor the operation of electrical appliances and hence keep a balance between energy consumption and users' comfort. In this regard, several IoT control techniques have been proposed in the literature to manage energy consumption based on proportional-integral-derivative (PID) control, Fuzzy control, ANNs, predictive control, dynamic models and optimization techniques.
\begin{itemize}
\item \textbf{G1. PID control:} it is a well-known control algorithm used for energy saving and home automation. It is based on the utilization of a control loop approach  using feedback. It is used to automatically monitor appliance power consumption, minimize energy usage of heating/cooling systems and guarantee good indoor environmental quality \cite{Blasco2011,ULPIANI20161,Wang2018sage}. 
\item \textbf{G2. Fuzzy control:} is a control algorithm that uses fuzzy-logic mathematical systems to analyze collected data from various sources such as power consumption, temperature, humidity, luminosity and occupancy, and then monitor the appliances connected to the energy efficiency system to reduce wasted energy \cite{He7793706,GHADI2014290}. Fuzzy control has been introduced to overcome the limitations of the PID control \cite{ULPIANI20161}.
\item \textbf{G3. Artificial neural networks (ANNs):} a large number of IoT-based energy efficiency systems have adopted ANNs as the principal element to optimize the monitoring of electrical appliances and reduce power consumption \cite{Qolomany8754678,ZEKICSUSAC2020102074}. Therefore, ANNs are usually trained with reference to historical power consumption data describing the consumption behavior of a specific building and then efficient rules are generated to monitor various IoT devices \cite{Runge2019,Kalogirou2006,Ateeq2019}.
\item \textbf{G4. Predictive control:} is an advanced approach of energy consumption control that is used to monitor energy consumption of appliances and satisfy an ensemble of constraints \cite{SIROKY20113079,PREGLEJ2014520,app8030408}. It is mainly based on the use of linear empirical models. 
\item \textbf{G5. Optimization techniques:} they focus in minimizing power consumption by systematically using different kinds of data collected from smart meters and smart sensors installed in buildings \cite{Gupta7578889,Petri2017,KHEIRI2018897}. Consequently, different frameworks have utilized these techniques to optimize energy usage and reduce wasted energy, such as the particle swarm optimization (PSO) \cite{DELGARM2016293,Ullah2017}, genetic algorithms (GA) \cite{fenrg.2018.00025} and multi-objective GA (MOGA) \cite{Yu2015}.  
\item \textbf{G6. Dynamic models:} are simplified representations of some real-world entities. They are intended to mimic some essential features of the energy saving system while leaving out the inessential ones through describing how the power consumption changes over time. They usually employ logical flow diagrams assuming objectives, time, and cost along with decision-making criteria to model an energy saving system \cite{MOKHATAB2019579}. In this regard, they are essentially deployed to predict the evolution of energy consumption based on the adoption of recommended control actions \cite{Genco2015,FOUCQUIER2013272,doi:10.1177/1744259113475543}.
\end{itemize}

\begin{figure}[t!]
\begin{center}
\includegraphics[width=16cm, height=17.5cm]{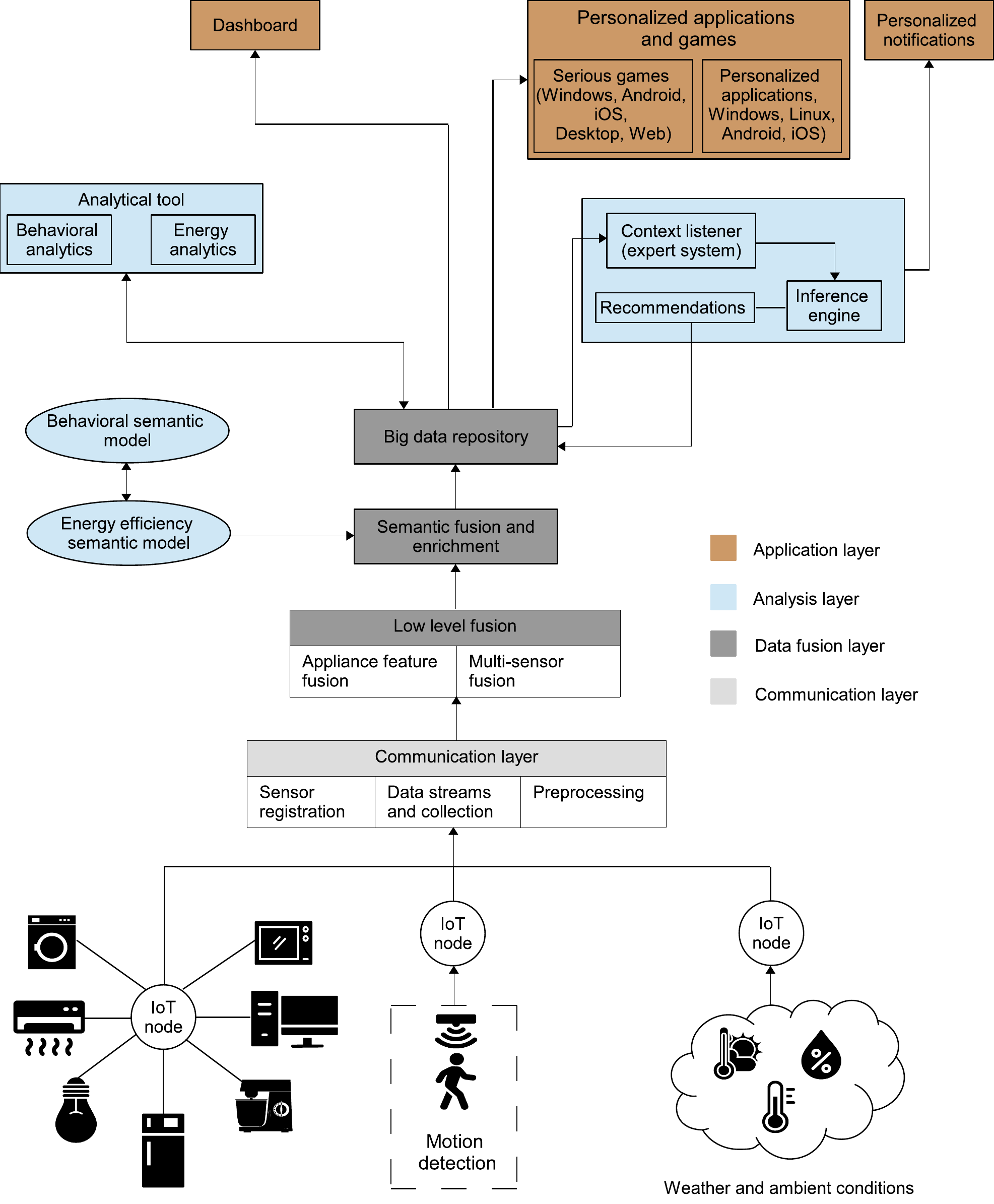}
\end{center}
\caption{Typical architecture of an energy efficiency system using information fusion strategies.}
\label{fig:flowchart0}
\end{figure}

Fig. \ref{fig:flowchart0} depicts a typical architecture of an energy efficiency system mainly including four layers called data fusion, communication, analysis and application layers. The communication layer is responsible for sensors registration, data collection and processing. Following, the fusion layer focuses on appliance feature fusion for appliance identification, multi-sensor fusion, semantic enrichment and data storage. Moving forward, the analysis layer refers to using different data mining, analytical tools and recommender systems to generate personalized advices. Finally, the application layer, that comes on the top of the analysis layer, aims mainly at developing mobile applications, social games and visualization tools to interact with end-users. Consequently, via adopting such architecture, a group of built-in and interconnected interfaces can be provided to endorse developing and provisioning of customized services, smartphone applications and serious games, which in turn attempt to positively influence energy consumption behavior of end-users.

\subsection{Comparison of energy efficiency projects} \label{EngFrameworks}

This section deals with the conducted works on data fusion based energy efficiency systems, most of which are related to human behavioral change. They describe different energy efficiency frameworks for persuading consumers and their corresponding methodologies. Most of the following frameworks are part of an initiative started by the Europe Commission under the Horizon 2020 program to reach \enquote{Secure, Clean and Efficient Energy} state \cite{H2020}. Thus, it is best to review what they have achieved so far as it is the most recent works done on energy efficiency.


In \cite{BIGFUSE2020}, BIGFUSE attempts to develop semantic-based information fusion and analysis techniques for (a) combining multimodality data from multiple sources; and (b) extracting practical knowledge to support the users to make the right decisions to reduce energy wastage. In \cite{Iadanza2014}, STREAMER architecture is proposed aiming at developing a semantic data fusion strategy to motivate buildings' occupants towards adopting energy efficient lifestyles. To achieve this challenge, this project relies on the fusion of data through the combination of building information models (BIM) and geographic information systems (GIS). 
In \cite{Garbi2019}, BENEFFICE  platform uses a multi-sensor data fusion strategy to monitor energy consumption in typical households and hence aids in reducing electricity usage via following energy saving recommendations triggered by a mobile application. The latter allows users earning CO$_{2}$ coins to exchange with real currency (Euros).

In \cite{ORBEET2016}, the ORBEET platform focuses on fusing data of BIM, business process models and real-time power consumption footprints. This was made possible by using a cloud computing server aiming at enabling electricity usage monitoring and establishing a real-time countability of individuals inside buildings and work spaces to a better energy management.
In \cite{Barthelmes2019a}, MOBISTYLE incorporates what other projects missed to add, which includes the health aspect of the energy consumers. This framework opted not only for providing guidance for reducing energy consumption, but also took into consideration suggestions for creating a healthy environment. Multiple sensors are used to measure indoor conditions such as temperature, humidity and CO$_{2}$ levels along with power consumptions, in addition to user occupancy, thermostat adjustments, windows and doors openings and outdoor environmental variables from online sources \cite{Tisov2017a}.

Papaioannou et al. \cite{Papaioannou2017a} introduce ChArGED ecosystem, in which the objective of creating a serious game is to persuade consumers to reduce energy consumption at peak hours. Based on given peak hours slots by the electric utility, contests between the players are designed to monitor their overall consumption using data fusion from multiple sources \cite{Papaioannou2017b}. In \cite{Luca2017}, GreenPlay architecture is proposed, which aims at raising consciousness among household users via implementing a real-time power usage monitoring platform along with developing a serious game. In \cite{Brudermann2015}, Brudermann et al. propose TRIBE project that targets how to contribute to an end-user behavioral change to achieve a high level energy saving in public buildings. This is done through (a) collecting and aggregating data from various sources; (b) engaging consumers in the adventure of playing a serious game that is connected to real-time data collection; and (c) providing actionable feedback to reduce wasted energy.
In \cite{Barbosa2017a}, the EnerGAware platform essentially focuses on providing a serious game heavily backed by IoT sensors fusion and integration using a service management system based on publish/subscribe approach. The main purpose of the serious game is to improve the likelihood of behavioral change for energy consumers via education and training. Rewards given in the serious game are based on improvements in their recorded energy consumption reductions. Furthermore, game progress is recorded to see the effect of the game with respect to real-life energy savings (i.e. how much the end-users have gathered points) \cite{Casals2017a}.
In \cite{Azarova2018}, the PEAKapp framework is proposed to incentivize consumers comport in an energy saving manner by supplying personalized saving pieces of advice through a social game application. Overall, data are collected and fused from different sensors to automatically provide discounts during peaks of power consumption for electricity produced from renewable energies. This results in significantly reducing carbon footprints.

In \cite{Fraternali2017a}, enCOMPASS architecture creates a recommendation system that provides energy saving suggestions based on the fusion of contextual data. Advised actions (recommendations) are generated based on collecting, fusing and analyzing data from various sources, including electricity consumption, user presence, indoor and outdoor temperatures and humidity. In addition,  the data visualization and user participation tool MyEnCompass \cite{Bernaschina2019a} along with the game application Funergy, are implemented to raise awareness regarding energy consumption, especially for kids \cite{Fraternali2018a}.
In \cite{Fotopoulou2017c}, ENTROPY strives to boost behavioral change towards energy saving in buildings environment. Interestingly, it uses a set of novel ICT elements, generally categorized by data fusion, semantic web, context-based recommendations, information retrieval and analysis strategies. Along with that, designs of both recommender system and serious game are established \cite{Garcia2017b}. In addition to data collected from power meters, ambient sensors and outside weather conditions, information related to the user such as his/her nickname, role, gender, age and score are also gathered.
In \cite{MYLONAS201943,Mylonas8628297}, the overall objective of the GAIA ecosystem is to reduce energy consumption of academic community, faculty, staff, students and parents. A recommender system is deployed to produce tailored recommendations based on the collection, fusion and analysis of multiple data sources from academic buildings. Towards this regard, an inventive ICT framework is implemented using web-based, mobile, social and sensing devices. 
In \cite{Apostolou2016}, the GreenSoul platform is deployed with the view of saving more than 20\% of energy consumption through behavioural modification recommendations. The latter are generated based on gathering and fusing data from several pilot buildings in different climate areas. Further, it can also properly evaluate the improvements made by end-users.

In \cite{BESTEnergy2013}, BestEenergy ecosystem uses a set of data visualization tools such as interactive websites and giant screens to inform end-users about their power usage footprints. To that end, various energy consumption fingerprints from different sources are recorded, fused and then visualized in order to come up with an energy saving decision-making feedback.   
While, in \cite{Korte2013}, eSESH platform provides efficient energy usage monitoring and control of electricity consumption mainly through visualizing and comparing consumption footprints collected and fused from a social housing network. Consequently, end-users could compare their consumption rates over short or long periods and can have access to a prediction of their energy usage. 
Following, SmartBuild framework \cite{Kyritsis2014} provides the end-users with visualization and statistical tools about the social-economic benefits that can be achieved when they follow a set of energy efficiency orientations. The latter are generated based on collecting, fusing and analyzing consumption fingerprints and comfort preferences from 9 pilot buildings. It also offers consumers the possibility to remotely monitor their electrical appliances and communicate their feedback about building comfort and satisfaction. 
Next, in \cite{SaveWork2017}, SAVE@WORK platform is developed in order to endorse energy saving in public buildings. This is mainly assisted with the implementation of online data visualization tools built upon data collection and fusion of various consumption and ambient conditions. Visualization provides end-users with real-time consumption statistics and comparisons to prompt them to shortage wasted energy.   

Finally, the EM3 initiative \cite{ALSALEMI2019ae,Sardianos2020iciot} is proposed, which allows exploiting power consumption data and ambient conditions to develop energy efficiency models and smart data analytics that highlight opportunities for consumers and businesses. This was possible through (a) data collection and fusion at different levels \cite{Himeur2020}; (b) data analysis using the micro-moment paradigm, which helps in detecting anomalous power consumption, such as excessive consumption and/or consumption while the end-user is outside; (c) a mobile recommender system that is used to provide end-users with personalized advices to optimize their consumption levels; and (d) novel interactive visualizations that allow end-users to easily comprehend their consumption footprints \cite{SardinosIEEEBigdata2020}.
Table \ref{EEcomp}, summarizes a list of several data fusion based energy efficiency ecosystems. It presents a comparison between these frameworks based on the different aspects discussed in the aforementioned sub-sections, including (1) data fusion level; (2) data fusion techniques; (3) behavioral change influencer; (4) behavioral change incentive; (5) recorded data; (6) platform architecture; (7) IoT technologies; and (8) application scenario.

\begin{table} [t!]
\caption{Comparison of data fusion based energy efficiency ecosystems.}
\label{EEcomp}
\begin{center}

\begin{tabular}{lllllllll}
\hline
{\scriptsize \textbf{Framework}} & {\scriptsize \textbf{Data fusion}} & 
{\scriptsize \textbf{Data fusion}} & {\scriptsize \textbf{Influencer}} & 
{\scriptsize \textbf{Incentive}} & {\scriptsize \textbf{Recorded data}} & 
{\scriptsize \textbf{Platform}} & {\scriptsize \textbf{IoT}} 
& {\scriptsize \textbf{Application scenario}} \\ 
& {\scriptsize \textbf{level}} & {\scriptsize \textbf{technique}} &  &  &  & 
{\scriptsize \textbf{architecture}} & {\scriptsize \textbf{technologies}} &  \\ \hline
{\scriptsize W1: BIGFUSE} & {\scriptsize L2} & {\scriptsize F1} & 
{\scriptsize I1, I3} & {\scriptsize C2} & {\scriptsize D1, D4, D5} & 
{\scriptsize A2} & {\scriptsize T1, T2, T4} & {\scriptsize %
Public buildings} \\ 
{\scriptsize W2: STREAMER} & {\scriptsize L3} & {\scriptsize F1, F8} & 
{\scriptsize I3} & {\scriptsize C1, C2} & {\scriptsize D1} & {\scriptsize A3}
& {\scriptsize T1, T4} & {\scriptsize Healthcare districts}
\\ 
{\scriptsize W3: BENEFFICE} & {\scriptsize L2} & {\scriptsize F8} & 
{\scriptsize I1} & {\scriptsize C4} & {\scriptsize D1} & {\scriptsize A3} & 
{\scriptsize T1, T4} & {\scriptsize Households} \\ 
{\scriptsize W4: ORBEET} & {\scriptsize L3} & {\scriptsize F1} & 
{\scriptsize I1, I3} & {\scriptsize C2} & {\scriptsize D1} & {\scriptsize A3}
& {\scriptsize T1, T4} & {\scriptsize Work spaces} \\ 
{\scriptsize W5: MOBISTYLE} & {\scriptsize L2} & {\scriptsize F1, F3, F4} & 
{\scriptsize I1} & {\scriptsize C2} & {\scriptsize D1, D2, D3} & 
{\scriptsize A4} & {\scriptsize T1, T2, T3, T4} & 
{\scriptsize Households} \\ 
{\scriptsize W6: ChArGED} & {\scriptsize L1, L2} & {\scriptsize F1, F2} & 
{\scriptsize I2} & {\scriptsize C1, C2, C3} & {\scriptsize D1} & 
{\scriptsize A2} & {\scriptsize T1, T4} & {\scriptsize %
Households} \\ 
{\scriptsize W7: GreenPlay} & {\scriptsize L2, L3} & {\scriptsize F1, F6} & 
{\scriptsize I2} & {\scriptsize C2, C3} & {\scriptsize D1} & {\scriptsize A1}
& {\scriptsize T1, T4} & {\scriptsize Households} \\ 
{\scriptsize W8: TRIBE} & {\scriptsize L2} & {\scriptsize F1, F7} & 
{\scriptsize I2} & {\scriptsize C2, C3} & {\scriptsize D1, D2} & 
{\scriptsize A3} & {\scriptsize T1, T2, T3, T4} & 
{\scriptsize Public buildings} \\ 
{\scriptsize W9: EnerGAware} & {\scriptsize L2} & {\scriptsize F1, F8} & 
{\scriptsize I2} & {\scriptsize C2, C3} & {\scriptsize D1, D2, D3} & 
{\scriptsize A4} & {\scriptsize T1, T2, T4} & {\scriptsize %
Households} \\ 
{\scriptsize W10: PEAKapp} & {\scriptsize L2} & {\scriptsize F1, F4} & 
{\scriptsize I1, I2} & {\scriptsize C1} & {\scriptsize D1} & {\scriptsize A4}
& {\scriptsize T1, T4} & {\scriptsize Households} \\ 
{\scriptsize W11: enCOMPASS} & {\scriptsize L2, L3} & {\scriptsize F1, } & 
{\scriptsize I1, I3} & {\scriptsize C1, C2} & {\scriptsize D1, D2} & 
{\scriptsize A2} & {\scriptsize T1, T2, T4} & {\scriptsize %
Households} \\ 
{\scriptsize W12: ENTROPY} & {\scriptsize L2} & {\scriptsize F1, F5} & 
{\scriptsize I2} & {\scriptsize C2, C3} & {\scriptsize D1, D2, D3} & 
{\scriptsize A4} & {\scriptsize T1, T2, T4} & {\scriptsize %
Households} \\ 
{\scriptsize W13: GAIA} & {\scriptsize L2} & {\scriptsize F1, F7} & 
{\scriptsize I2} & {\scriptsize C2, C3} & {\scriptsize D1, D4, D5} & 
{\scriptsize A3} & {\scriptsize T1, T2, T3} & {\scriptsize %
Academic buildings} \\ 
{\scriptsize W14: GreenSoul} & {\scriptsize L2} & {\scriptsize F1, F3} & 
{\scriptsize I1} & {\scriptsize C2} & {\scriptsize D1} & {\scriptsize A4} & 
{\scriptsize T1, T3} & {\scriptsize Public buildings} \\ 
{\scriptsize W15: BESTEnergy} & {\scriptsize L2} & {\scriptsize F1, F4} & 
{\scriptsize I3} & {\scriptsize C1, C2} & {\scriptsize D1} & {\scriptsize A4}
& {\scriptsize T1, T3} & {\scriptsize Work spaces} \\ 
{\scriptsize W16: eSESH} & {\scriptsize L2} & {\scriptsize F4} & 
{\scriptsize I3} & {\scriptsize C2} & {\scriptsize D1} & {\scriptsize A3} & 
{\scriptsize T1, T3} & {\scriptsize Households} \\ 
{\scriptsize W17: SmartBuild} & {\scriptsize L2} & {\scriptsize F1, F4} & 
{\scriptsize I3} & {\scriptsize C1, C2} & {\scriptsize D1, D2, D4, D5} & 
{\scriptsize A3} & {\scriptsize T1, T2, T3, T4} & 
{\scriptsize Households} \\ 
{\scriptsize W18: SAVE@WORK} & {\scriptsize L2} & {\scriptsize F1} & 
{\scriptsize I3} & {\scriptsize C1, C1} & {\scriptsize D1} & {\scriptsize A3}
& {\scriptsize T1, T2, T3} & {\scriptsize Work spaces} \\ 
{\scriptsize W19: EM3} & {\scriptsize L1, L2} & {\scriptsize F1,F3, F5,F6,F8}
& {\scriptsize I1, I4} & {\scriptsize C2} & {\scriptsize D1, D2, D4} & 
{\scriptsize A1} & {\scriptsize T1, T2, T4} & {\scriptsize %
Academic buildings} \\ 
&  &  &  &  &  &  &  & {\scriptsize and households}  \\ \hline
\end{tabular}

\end{center}
\end{table}


\subsection{Discussion and important findings}
Under this framework, different data fusion based energy efficiency systems have been described, reviewed and classified according to various parameters. In what follows, we derive the important findings based on what has been discussed in the previous lines. This helps in appropriately guiding us to map future orientations for improving energy efficiency systems. Accordingly, a set of relevant findings has been drawn as follows: 
\begin{itemize}
\item Although visualization tools play an essential role to motivate end-users promoting energy saving behaviors via a better understanding of energy consumption footprints, most of the energy efficiency frameworks fail to implement them along with other incentive tools. The EM3 is among the first frameworks to implement visualization tools in combination with recommender systems, and hence help convincing end-users to replace inefficient energy habits with efficient ones.
\item Most of the studied energy efficiency frameworks adopt a sensor level fusion, this is because of the increasing use of IoT smart sensors to collect energy consumption and contextual information. The appliance level fusion has been considered in few frameworks, such as ChArGED and EM3. This kind of fusion helps effectively in reducing the installation cost of energy efficiency solutions since appliance-level consumption data are inferred directly from the main record and no extra installation is required.
\item Although some energy frameworks have already started considering the collection of data from different modalities, a large number of these frameworks have only gleaned energy consumption patterns and ignored other data sources, which have a significant impact on increasing energy usage.
\item Cloud computing architectures have been adopted in a large number of energy efficiency frameworks, while few of them turn to the use of fog computing architectures, which allow a low-cost implementation, among them, EM3 and GreenPlay frameworks. 
\item Data association is the most deployed fusion technique because its concept is very simple and relies on the combination of various types of data. Moreover, an increasing interest has been paid to other fusion strategies, including the statistical inference and visualization. In addition, energy efficiency in EM3 is achieved through the use of up to five fusion techniques.  
\item MOBISTYLE and TRIBE are among the few frameworks, that in addition to providing end-users with personalized advices to reduce their energy consumption, they allow them to monitor the usage of electrical appliances and equipment automatically via adopting appropriate actuators.
\end{itemize}

On the other side, as discussed previously, information fusion in each energy efficiency ecosystem has its own importance and entails the combination of various kinds of data. Thus, aiming at providing a more thorough assessment, the following points should be underlined:
\begin{itemize}
\item Some energy efficiency ecosystems concentrate on collecting one kind of data, i.e. power consumption footprints. This kind of systems with unique modality data can be developed easily, however, it is not effective because they do not take into consideration other parameters influencing energy usage. In other words, multi-modality data help greatly in designing more powerful energy saving solutions.
\item Some energy efficiency frameworks focus on achieving energy savings from the end-user's perspective via analyzing his/her behavior. However, end-user personal data and household-related information are private, and hence they require to be protected. Data transmitted using novel IoT communication solutions may be less sensitive in terms of privacy if adequate encryption and privacy protection solutions are adopted.
\item In addition to the fact that data fusion in energy efficiency systems should be reliable, the quality of collected data plays a major role in improving their reliability and performance. Therefore, because of the high distribution of smart meters and smart sensors, the use of powerful protocols ensuring a good quality of data collection and transmission becomes an ultimate priority.
\end{itemize}  

\section{Data fusion for appliance identification} \label{sec3}
This section presents an example of the application of data fusion strategies in the EM3 energy efficiency framework. Accordingly, a novel appliance identification system based on the fusion of appliance features is proposed. It aims at exploring the 2D descriptors and information fusion to construct robust appliance features.

\subsection{Background/concept of 2D descriptors}
Recently, 2D local descriptors have aroused a growing interest in different research areas, such as image processing and computer vision \cite{Himeur2018MTAP}, fault detection \cite{Khan2016}, medical applications \cite{NANNI2010117,Tripathi7848388} and face recognition \cite{Kas2019}. They are usually used to extract pertinent features after splitting the whole 2D signal into several small regions using square patches. Specifically, 2D local descriptors can be calculated at each key point of the 2D representation to derive further details about the region neighboring that key point. Extracted characteristics are then combined into a unique, spatially enhanced feature vector effectively describing the initial 2D signal.

2D descriptors can be classified based on the data type deployed to describe their feature vector components. Accordingly, if a unique bit per component is utilized, the descriptor is considered as binary, otherwise, it is identified as non-binary. The first category is generally less robust, but faster, less complex and more
compact than the second category \cite{Bellavia8757967}. Additionally, 2D descriptors are also classified based on whether they utilize a priori data knowledge and if they help in training ML models. If that is true, a descriptor is labeled as data-driven, and hand-crafted otherwise. Moreover, a 2D descriptor usually adopts different kinds of information to encode the initial 2D signals through a small region (square or circle), such as the intensity, gradient, texture or gray level \cite{Krig2014}.

\subsection{2D descriptors for appliance identification}
\begin{figure}[t!]
\begin{center}
\includegraphics[width=15cm, height=2.6cm]{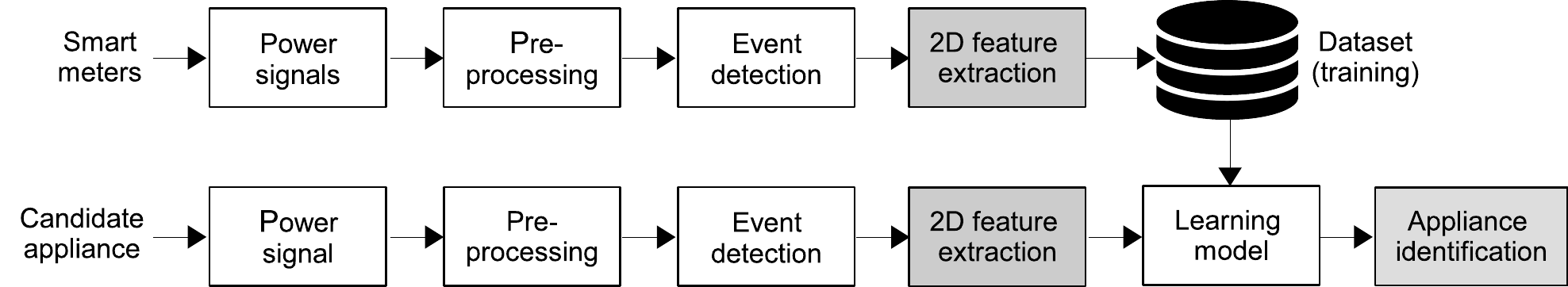}
\end{center}
\caption{Block diagram of the proposed appliance identification system.}
\label{BlockDiagram}
\end{figure}

In this framework, the 2D descriptors are used to extract pertinent appliance features from power consumption signals. To the best of our knowledge, this is the first time that this kind of descriptor is used to extract appliance fingerprints. To allow this extraction, 1D power signals are firstly transformed to the 2D space. Following, 2D local descriptors are applied to extract short histograms and their outputs are then fused. This results in efficient feature representations that are finally used to classify appliances. Fig. \ref{BlockDiagram} represents the block diagram of the appliance identification system used in our EM3 energy efficiency framework.

The pre-processing stage refers to cleaning collected raw power signals by checking for missing observations and thus replacing each zero value by the average of its neighbor. Further, because the essential objective of this framework relies on designing a novel feature extraction scheme based on the fusion of 2D descriptors, in the event detection step, the NILMTK module \cite{Batra2014ACM} has been used to detect the events in each power signal.

To design appliance identification systems, many papers have focused on developing event detection and 1D feature extraction approaches. However, extracting features using 1D local descriptors is fairly ineffective because these kind of descriptors only rely on extracting data from one 1D blocks, which do not provide pertinent information to efficiently encode power signals. In contrast, 2D local descriptors have proved their efficiency in many fields such as face recognition, indexing and retrieval of medical images, iris identification, video content analysis and even Electrocardiogram (ECG) classification. Driven by the efficacy of 2D local descriptors, we present in this section a powerful appliance identification approach based on the fusion of Weber local descriptor (WLD) and local binary pattern (LBP). In addition, we experiment with several ML classifiers to find out the best model.

\subsection{Fusion of 2D descriptors}
\begin{figure}[t!]
\begin{center}
\includegraphics[width=16.5cm, height=10.1cm]{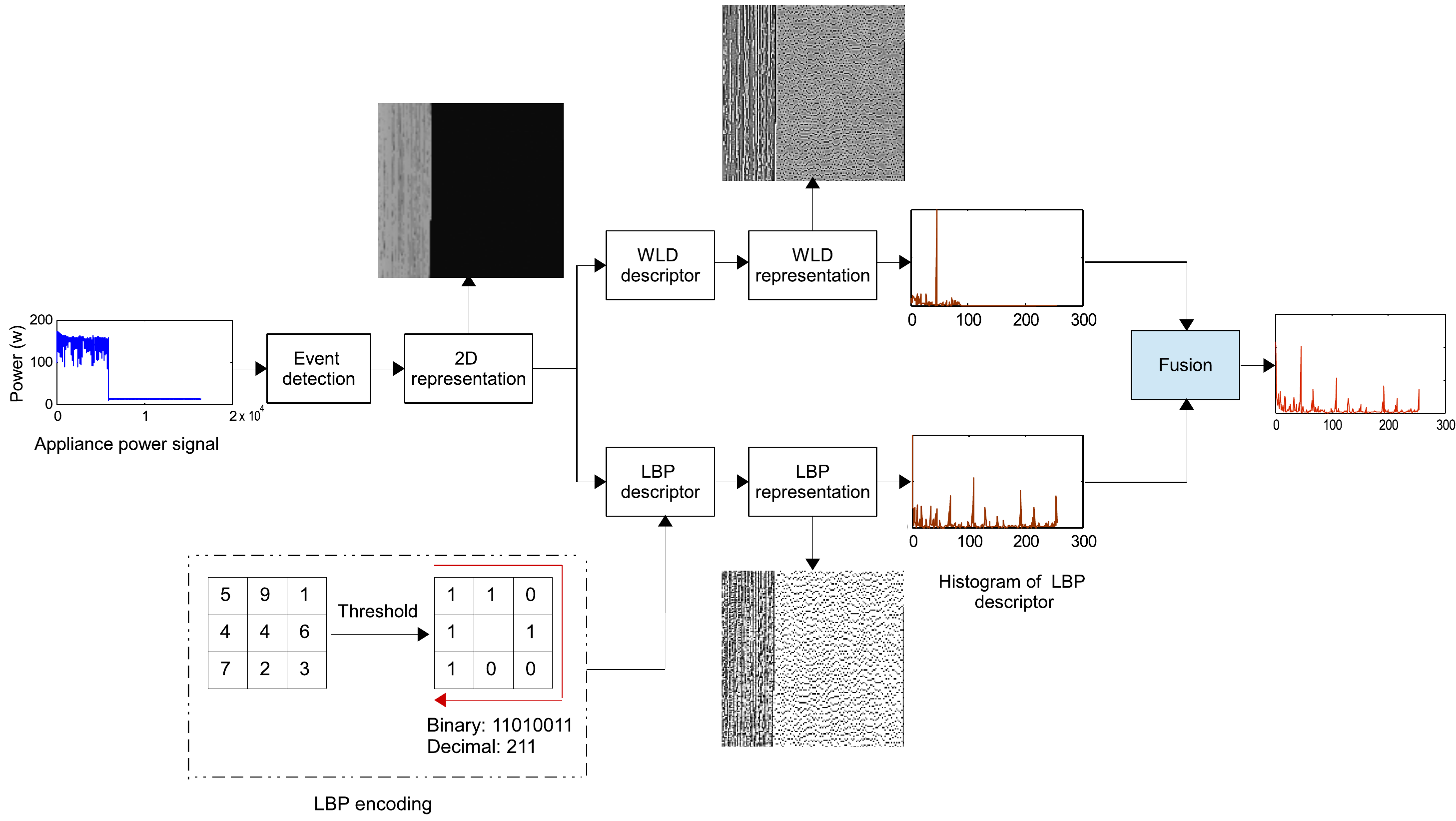}
\end{center}
\caption{Flowchart of the proposed feature descriptor based on the fusion of 2D descriptors.}
\label{FlowFusion}
\end{figure}

The flowchart in Fig. \ref{FlowFusion} explains the proposed feature extraction algorithm based on the fusion of 2D descriptors used in implementing the appliance identification system. This solution is based on simple yet effective fusion using the summation of appliance features extracted separately from WLD and LBP descriptors.

\subsubsection{Weber local descriptor (WLD)}
\begin{figure}[t!]
\begin{center}
\includegraphics[width=2.8cm, height=2.8cm]{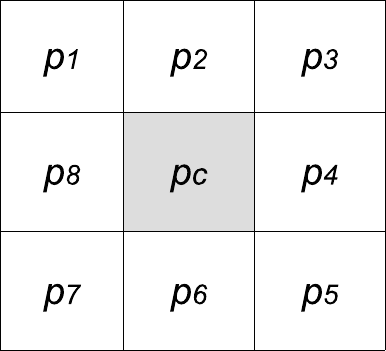}
\end{center}
\caption{$3 \times 3$  neighborhood of the power sample $p_{c}$.}
\label{patchFig}
\end{figure}

WLD is a powerful descriptor that has been proposed initially to extract local texture features from images. It is mainly built extracting differential excitations and gradient orientations. It has been inspired by psychological law called Weber's Law \cite{Dawood2014}.
After transforming the 1D power signal into a 2D space, the differential
excitation is computed through estimating the ratio between the sum of power
magnitude variations of the central power sample versus its neighbor's power
samples over the magnitude of the central power sample. This is achieved
using the following equation \cite{Chen5204092}:

\begin{equation}
\zeta (p_{c})=\arctan \left[ \sum\limits_{i=1}^{N}\left( \frac{p_{i}-p_{c}}{%
p_{c}}\right) \right] 
\end{equation}

where $p_{c}$ represents the central power sample, $\zeta (p_{c})$ refers to
the differential excitation of the central power sample, $N$ is the number
of neighboring samples ($N=8$), and $p_{i}$ stands for the $i$th sample
neighbor of $p_{c}$. Moreover, the gradient orientation is then computed \
based on Eq. \ref{eq2}:

\begin{equation}
\theta =\arctan \left( \frac{o_{10}}{o_{11}}\right) 
\label{eq2}
\end{equation}

where $o_{10}=p_{6}-p_{2}$ and $o_{11}=p_{8}-p_{4}$. Where $p_{2},$ $p_{4},$ 
$p_{6}$ and $p_{8}$ are the neighboring samples of $p_{c}$ in a patch of $%
3\times 3$ as illustrated in Fig. \ref{patchFig}.

Following, a 1D histogram of the gradient orientation is extracted as explained in \cite{Chen5204092} to represent each power signal.

\subsubsection{Local binary pattern (LBP)}
LBP descriptor is deployed to abstract a histogram representation of the power signal values from their 2D representation. It mainly focuses on encoding the local structure around a central power sample in each specific block \cite{Huang5739539}. As illustrated in Fig. \ref{FlowFusion}, every central sample is compared with its surrounding patterns in a patch of $3 \times 3$ neighbors via the subtraction of the central sample value. An encoding process follows in which positive values become 1 while negative values are shifted to 0. A binary stream is then collected using a clockwise strategy. The extracted binary patterns are the corresponding LBP codes. The next step is to collect all the binary codes from all the kernels in a binary 2D representation that is transformed to a decimal representation. Finally, a histogramming approach is processed on the resulting decimal matrix to extract the histogram of the LBP representation. The proposed feature extraction process is outlined in Algorithm \ref{algo1} \cite{Ahonen1717463}. 

\begin{algorithm}[ht!]
\SetAlgoLined

\KwResult{$\mathrm{H}_{LBP}$: The histogram of local binary patterns (LBP) }

Set the power signals matrix $Y(i,j)$, where $i$ is the index of feature vectors and $j$ is the index of patterns in each vector;

 \While{$i \leq M$ (\textnormal{with} $M$ \textnormal{being the total number of power signals in the whole dataset})}{

1. Normalize and convert the power signal $i$ into a 2D representation.

2. Estimate the LBP values for a power sample $(u_{c},v_{c})$ in the power matrix using an $S \times S$ patch as follows
\begin{equation}
LBP_{n,S}(u_{c},v_{c})=\sum\limits_{n=1}^{N}b(j_{n}-j_{c})2^{n}
\end{equation}
where $j_{c}$ and $j_{n}$ represent the power value of the central sample and $n$ the surrounding power values in the square neighborhood with the patch size $S$. The binary encoding function $b(u)$ can be given as:
\begin{equation}
b(u)=\left\{ 
\begin{array}{cc}
1 & \mathrm{if}~u \geq 0 \\ 
0 & \mathrm{if}~u < 0%
\end{array}%
\right. 
\end{equation}

3. Collect the binary patterns $LBP_{n,S}(u_{c},v_{c})$ from each block and convert the obtained binary matrix to a decimal representation.

4. Apply a histogramming process on the novel decimal matrix to derive the histogram of LBP representation $H_{LBP}(n,S)$ estimated through each block and use it as texture features to identify electrical devices based on their 2D representations. The descriptor $\mathrm{H}_{LBP}(n,S)$ produces $2^{n}$ new values related to $2^{n}$ binary samples generated by $N$ neighboring power samples of each patch. 
}

\caption{The LBP algorithm applied on the power consumption signals to extract the LBP histograms.}
\label{algo1}
\end{algorithm}

\subsection{Fusion procedure}
After detecting the power event vector from a specific appliance power consumption signal, the obtained event vector is transformed into 2D space. Following, LBP and WLD descriptors are then applied on the 2D representation separately to derive two feature histograms, each one with a length of 256 bins. After that, LBP and WLD feature histograms are fused based on one of the three fusion strategies adopted in this framework, i.e. the summation, concatenation or multiplication. Specifically, the summation fusion is based on summing the LBP and WLD histograms. The concatenation fusion refers to concatenating both LBP and WLD histograms into a single histogram of length 512 bins. Finally, the multiplication fusion is based on multiplying LBP histogram by WLD histogram to come out with a novel histogram with similar length. The final fused histogram is then used as a unique representation for each specific appliance and this operation is repeated for all power signals collected from different appliances.

\subsection{Experimental results}
\subsubsection{Dataset description}

The performance of the proposed appliance identification scheme is assessed using three open access datasets, namely GREEND \cite{GREEND2014}, PLAID \cite{PLAID2014} and WHITED \cite{WHITED2016}, which record energy usage patterns at a device-level for various electrical appliances in different domestic buildings. Data are gathered at different sampling rates that are 1 Hz, 30 kHz and 44 kHz, respectively. The GREEND datatset includes six domestic appliance classes and for each class, data are observed at a daily-level for more than six months. Explicitly, each observation represents power consumption of a specific appliance during a whole day. In contrast, both the PLAID and WHITED datasets include power consumption footprints from 11 appliance categories and each category include different consumption observations for the same appliance type but collected from devices of different manufacturers. Table \ref{DataSets} summarizes the monitored appliance groups in each dataset for the performance assessment.

\begin{table}[t!]
\caption{Description of monitored appliances and their number for both the PLAID and WHITED datasets and observed days for the GREEND dataset.}
\label{DataSets}

\begin{center}

\begin{tabular}{lll|lll|lll}
\hline
\multicolumn{3}{c|}{\small PLAID} & \multicolumn{3}{|c|}{\small WHITED} & 
\multicolumn{3}{|c}{\small GREEND} \\ \hline
{\small Tag} & {\small Apliance} & {\small \# } & {\small Tag} & {\small %
Apliance} & {\small \# } & {\small Tag} & {\small Appliance} & {\small \# }
\\ 
& {\small category} & {\small app} &  & {\small category} & {\small app} & 
& {\small Category} & \multicolumn{1}{r}{\small days} \\ \hline
{\small 1} & {\small Fluorescent lamp} & {\small 90} & {\small 1} & {\small %
Modems/receivers} & \multicolumn{1}{r|}{\small 20} & {\small 1} & {\small %
Television} & \multicolumn{1}{r}{\small 258} \\ 
{\small 2} & {\small Fridge} & {\small 30} & {\small 2} & {\small Compact fluorescent lamp} & 
\multicolumn{1}{r|}{\small 20} & {\small 2} & {\small Network attached storage} & 
\multicolumn{1}{r}{\small 258} \\ 
{\small 3} & {\small Hairdryer} & {\small 96} & {\small 3} & {\small Charger}
& \multicolumn{1}{r|}{\small 30} & {\small 3} & {\small Washing machine} & 
\multicolumn{1}{r}{\small 238} \\ 
{\small 4} & {\small Microwave} & {\small 94} & {\small 4} & {\small Coffee
machine} & \multicolumn{1}{r|}{\small 20} & {\small 4} & {\small Dishwasher}
& \multicolumn{1}{r}{\small 257} \\ 
{\small 5} & {\small Air conditioner} & {\small 51} & {\small 5} & {\small %
Drilling machine} & \multicolumn{1}{r|}{\small 20} & {\small 5} & {\small %
Notebook} & \multicolumn{1}{r}{\small 258} \\ 
{\small 6} & {\small Laptop} & {\small 107} & {\small 6} & {\small Fan} & 
\multicolumn{1}{r|}{\small 30} & {\small 6} & {\small Coffee machine} & 
\multicolumn{1}{r}{\small 250} \\ 
{\small 7} & {\small Vacuum cleaner} & {\small 8} & {\small 7} & {\small Flat iron} & 
\multicolumn{1}{r|}{\small 20} &  &  &  \\ 
{\small 8} & {\small Incadescent light bulb} & {\small 79} & {\small 8} & 
{\small LED light} & \multicolumn{1}{r|}{\small 20} &  &  &  \\ 
{\small 9} & {\small Fan} & {\small 96} & {\small 9} & {\small Kettles} & 
\multicolumn{1}{r|}{\small 20} &  &  &  \\ 
{\small 10} & {\small Washing machine} & {\small 22} & {\small 10} & {\small %
Microwave} & \multicolumn{1}{r|}{\small 20} &  &  &  \\ 
{\small 11} & {\small Heater} & {\small 30} & {\small 11} & {\small Iron} & 
\multicolumn{1}{r|}{\small 20} &  &  &  \\ \hline
\end{tabular}

\end{center}
\end{table}

%

\subsubsection{Discussion}
Table \ref{ACCFscore-Comp} presents the accuracy and F1 score of the proposed appliance identification, in which various classifiers with different parameters are considered, including linear discriminant analysis (LDA), DT, deep neural networks (DNNs), EBT, SVM, KNN. The collected results are summed to fuse appliance features. It is worth mentioning that the KNN classifier with K=1/Euclidean distance outperforms the other ML algorithms under all datasets considered in this case study. Up to 99.68\% accuracy and 99.52\% F1 score are obtained with the GREEND dataset. Further, the proposed solution performs well on the WHITED repository, in which 99.06\% and 98.32\ F1 score are attained.

\begin{table}[t!]
\caption{The accuracy and F1 score results obtained using the proposed fusion of 2D descriptors with different classifiers.}
\label{ACCFscore-Comp}
\begin{center}

\begin{tabular}{lccccccc}
\hline
{\small Classifier } & {\small Classifier} & \multicolumn{2}{c}{\small GREEND%
} & \multicolumn{2}{c}{\small PLAID} & \multicolumn{2}{c}{\small WHITED} \\ 
\cline{3-8}\cline{3-7}
& {\small \ parameters} & {\small Accuracy} & {\small F1 score} & {\small %
Accuracy} & {\small F1 score} & {\small Accuracy} & {\small F1 score} \\ \hline
{\small LDA} & {\small /} & \multicolumn{1}{l}{\small 94.56} & 
\multicolumn{1}{l}{\small 94.49} & \multicolumn{1}{l}{\small 86.73} & 
\multicolumn{1}{l}{\small 83.67} & \multicolumn{1}{l}{\small 84.31} & 
\multicolumn{1}{l}{\small 83.66} \\ 
{\small DT} & {\small Fine, 100 splits} & \multicolumn{1}{l}{\small 93.98} & 
\multicolumn{1}{l}{\small 93.86} & \multicolumn{1}{l}{\small 84.9} & 
\multicolumn{1}{l}{\small 82.74} & \multicolumn{1}{l}{\small 93.85} & 
\multicolumn{1}{l}{\small 93.16} \\ 
{\small DT} & {\small Medium, 20 splits} & \multicolumn{1}{l}{\small 91.25}
& \multicolumn{1}{l}{\small 91.17} & \multicolumn{1}{l}{\small 74.82} & 
\multicolumn{1}{l}{\small 65.26} & \multicolumn{1}{l}{\small 93.53} & 
\multicolumn{1}{l}{\small 92.89} \\ 
{\small DT} & {\small Coarse, 4 splits} & \multicolumn{1}{l}{\small 76.64} & 
\multicolumn{1}{l}{\small 74.38} & \multicolumn{1}{l}{\small 59.43} & 
\multicolumn{1}{l}{\small 54.39} & \multicolumn{1}{l}{\small 50.28} & 
\multicolumn{1}{l}{\small 48.96} \\ 
{\small DNNs} & {\small 50 hidden layers} & \multicolumn{1}{l}{\small 74.76}
& \multicolumn{1}{l}{\small 73.1} & \multicolumn{1}{l}{\small 83.75} & 
\multicolumn{1}{l}{\small 80.09} & \multicolumn{1}{l}{\small 83.68} & 
\multicolumn{1}{l}{\small 83.47} \\ 
{\small EBT} & {\small 30 learners, 42 k splits} & \multicolumn{1}{l}{\small 90.31}
& \multicolumn{1}{l}{\small 89.76} & \multicolumn{1}{l}{\small 86.84} & 
\multicolumn{1}{l}{\small 83.65} & \multicolumn{1}{l}{\small 93.52} & 
\multicolumn{1}{l}{\small 92.37} \\ 
{\small SVM} & {\small Linear Kernel} & \multicolumn{1}{l}{\small 96.13} & 
\multicolumn{1}{l}{\small 96.49} & \multicolumn{1}{l}{\small 87.59} & 
\multicolumn{1}{l}{\small 84.78} & \multicolumn{1}{l}{\small 87.37} & 
\multicolumn{1}{l}{\small 85.96} \\ 
{\small SVM} & {\small \ Gaussian kernel} & \multicolumn{1}{l}{\small 93.78}
& \multicolumn{1}{l}{\small 92.93} & \multicolumn{1}{l}{\small 89.38} & 
\multicolumn{1}{l}{\small 85.75} & \multicolumn{1}{l}{\small 87.44} & 
\multicolumn{1}{l}{\small 88.29} \\ 
{\small SVM} & {\small Quadratic kernel} & \multicolumn{1}{l}{\small 95.17}
& \multicolumn{1}{l}{\small 94.96} & \multicolumn{1}{l}{\small 92.7} & 
\multicolumn{1}{l}{\small 90.61} & \multicolumn{1}{l}{\small 94.33} & 
\multicolumn{1}{l}{\small 92.06} \\ 
{\small KNN} & {\small K=10/Weighted Euclidean distance} & \multicolumn{1}{l}{\small 97.39} & 
\multicolumn{1}{l}{\small 97.11} & \multicolumn{1}{l}{\small 89.42} & 
\multicolumn{1}{l}{\small 83.36} & \multicolumn{1}{l}{\small 91.57} & 
\multicolumn{1}{l}{\small 88.75} \\ 
{\small KNN} & {\small K=10/Cosine distance} & \multicolumn{1}{l}{\small 97.51}
& \multicolumn{1}{l}{\small 97.35} & \multicolumn{1}{l}{\small 84.19} & 
\multicolumn{1}{l}{\small 81.6} & \multicolumn{1}{l}{\small 90.36} & 
\multicolumn{1}{l}{\small 91.36} \\ 
{\small KNN} & {\small K=1/Euclidean distance } & \multicolumn{1}{l}{{\small 
\textbf{99.68}}} & \multicolumn{1}{l}{{\small \textbf{99.52}}} & 
\multicolumn{1}{l}{{\small \textbf{96.73}}} & \multicolumn{1}{l}{{\small 
\textbf{96.43}}} & \multicolumn{1}{l}{{\small \textbf{99.06}}} & 
\multicolumn{1}{l}{{\small \textbf{98.32}}} \\ \hline
\end{tabular}

\end{center}
\end{table}

Table \ref{ACCFscore} illustrates a performance comparison of the proposed identification solution using a summation based fusion  in comparison with other two fusion techniques based on multiplication and concatenation. It is evident that the summation based fusion achieves better accuracy and F1 score under the three datasets used in this study. Moreover, the highest performance is obtained under the GREEND dataset, in which 99.68\% accuracy and 99.52\% F1 score are attained. However, the performance slightly drops under the PlAID dataset, where 96.73\% accuracy and 96.43\% F1 score are realized. This occurs because this dataset is highly imbalanced and the consumption fingerprints are collected for very short periods. Consequently, the consumption signatures of electrical devices are not perfectly captured, which leads to an increase in the misclassification rate.

\begin{table}[t!]
\caption{The accuracy and F1 score results obtained using different fusion strategies.}
\label{ACCFscore}
\begin{center}

\begin{tabular}{lllllll}
\hline
\textbf{Fusion method} & \multicolumn{2}{c}{\textbf{GREEND}} & 
\multicolumn{2}{c}{\textbf{PLAID}} & \multicolumn{2}{c}{\textbf{WHITED}} \\ 
& {\small accuracy} & {\small F1 score} & {\small accuracy} & {\small F1 score}
& {\small accuracy} & {\small F1 score} \\ \hline
{\small Multiplication} & 95.96 & 95.34 & 93.74 & 93.97 & 96.08 & 94.4 \\ 
{\small Concatenation} & 97.19 & 96.87 & 94.81 & 94.75 & 96.9 & 95.63 \\ 
{\small Summation} & \textbf{99.68} & \textbf{99.52} & \textbf{96.73} & 
\textbf{96.43} & \textbf{99.06} & \textbf{98.32} \\ \hline
\end{tabular}

\end{center}
\end{table}

\begin{figure}[t!]
\begin{center}
\includegraphics[width=8cm, height=5.2cm]{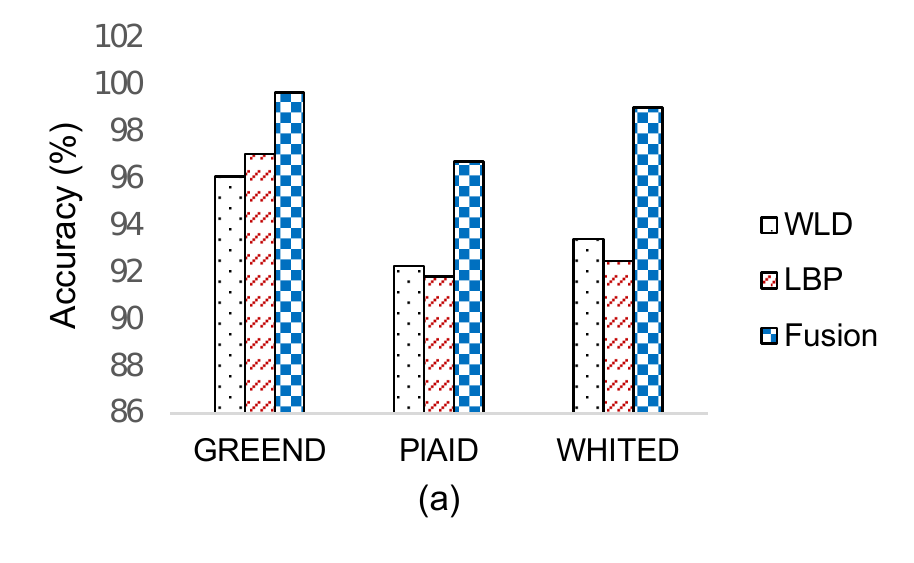}
\includegraphics[width=8cm, height=5.2cm]{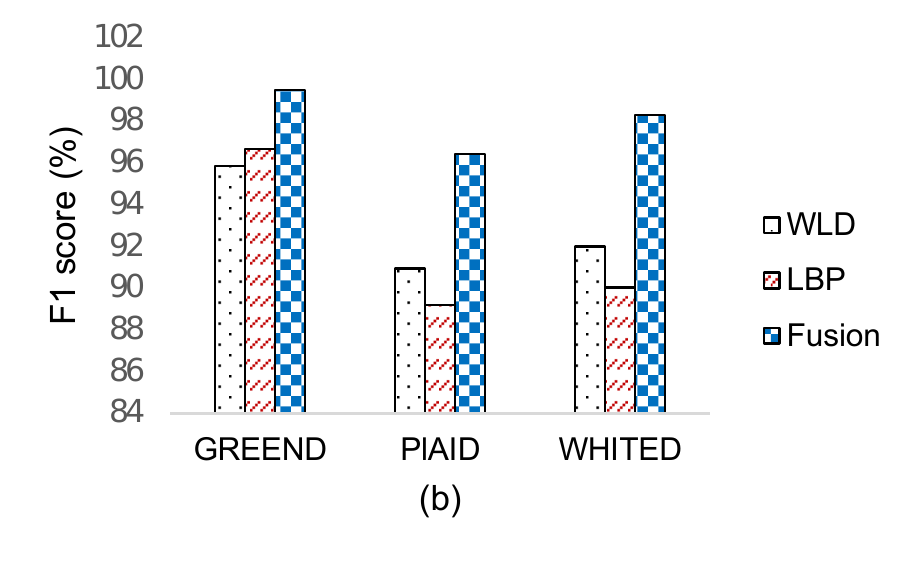}

\end{center}
\caption{Accuracy and F1 score of the proposed fusion technique with comparison to use of the WLD and LBP descriptors, separately.}
\label{ACCF1-comp}
\end{figure}

Fig. \ref{ACCF1-comp} depicts the success rates of 2D descriptors fusion, mainly under the KNN classifier (K=1/Euclidean distance), in comparison with the use of each descriptor separately. It is obvious that the highest performance has been obtained under the GREEND and WHITED dataset, while in the case of the PLAID dataset, the accuracy and F1 score have slightly dropped, this is mainly due to the imbalance property of this dataset, as explained previously.

\begin{figure}[t!]
\begin{center}
\includegraphics[width=15.3cm, height=2.8cm]{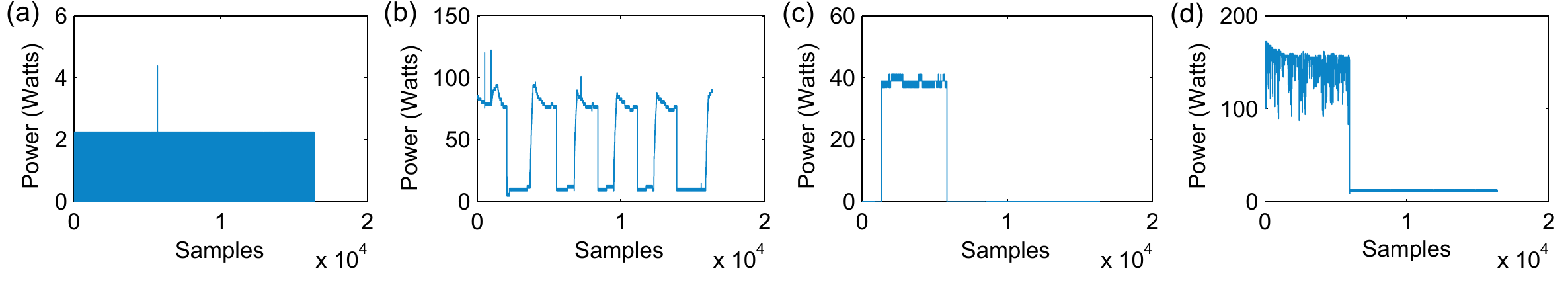}\\
(I) Example of power signals from the GREEND dataset: (a) Television , (b) Washing machine, (c) Notebook and (d) Coffee machine\\
\includegraphics[width=15.3cm, height=7.4cm]{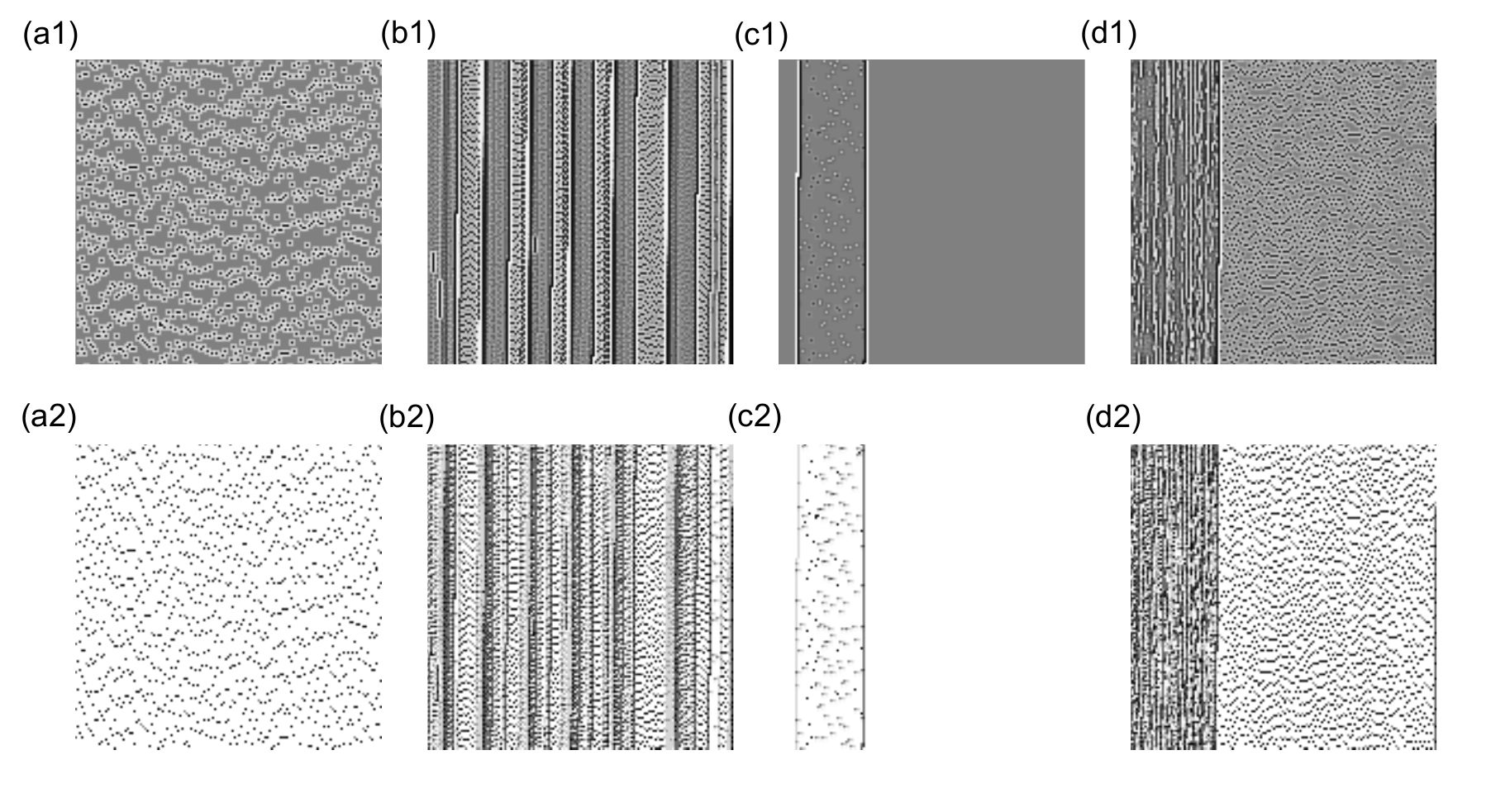}\\
(II) 2D representations of the power signals, in which (a1), (b1), (c1) and (d1) stand for WLD representations and (a2), (b2), (c2) and (d2) refer to LBP representations. \\
\includegraphics[width=15.3cm, height=5.8cm]{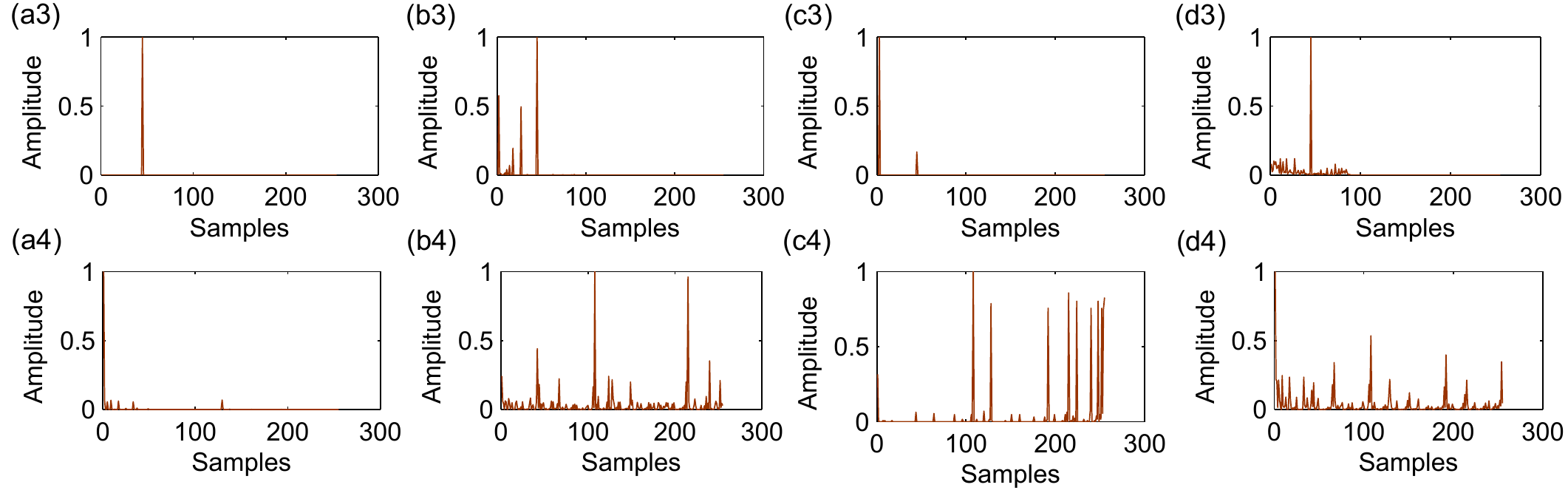}\\
(III) Histogram representations of the power signals, in which (a3), (b3), (c3) and (d4) stand for histograms of WLD representations and (a4), (b4), (c4) and (d4) refer to histograms of LBP representations. \\

\end{center}
\caption{Example of four appliance power signals from the GREEND dataset with their 2D and histogram representations, respectively.}
\label{rawSig-2D}
\end{figure}

Fig. \ref{rawSig-2D} demonstrates four appliance power signals, their 2D representations obtained using the WLD and LBP descriptors, and further their histograms extracted from the WLD and LBP spaces as well. It can be easily seen that 2D representations are quiet related to the shapes of 1D power signals and how the samples are distributed through the time axis. This can be seen in the case of the washing machine, notebook and coffee machine. Consequently, the 2D representation maintains some properties of the power signal and offers more possibilities to encode and extract features since each sample in the 2D space has 8 neighbors.

It is worth noting that the proposed appliance identification system based on the fusion of LBP and WLD descriptors uses a simple yet effective summation-based fusion strategy. The latter can be implemented without adding any complexity to the appliance identification task. Specifically, the only complexity of the proposed system is related to both LBP and WLD descriptors, which can be implemented at a very low complexity using Raspberry Pi or other low-cost platforms, as described in \cite{RPI3book,Baba2019}. LBP and WLD have been successfully applied for real-time applications using low-cost platforms for face recognition and video surveillance applications \cite{RPI3book}. Therefore, because power signals are less complex than images and videos, their computation complexity is less complex as well and their time execution is faster.

All in all, for the case of the EM3 framework that is deployed in Qatar, the proposed strategy is implemented using low-cost platforms, i.e. Raspberry Pi4 (RPI4) and/or Jetson Nano. RPI4 consumes up to 6.4 W/h when the module is fully utilized, while Jetson Nano consumes at most 10 W/h when the module is fully utilized. In addition, giving that the average yearly power consumption per capita in Qatar is 13388 kWh \cite{WorldData2020} and the average number of occupants per household is 5.5 \cite{MDPS2020}, a comparison is conducted to highlight the yearly power consumption per household, saved power, and power consumption of RPI4 and Jetson Nano. Further, the power consumption cost along with the saved cost are estimated as well, knowing that the 1 kWh cost is 0.032 US Dollars (USD).

Table \ref{PC-Comp} summarizes the results of this comparison. Therefore, it is clearly shown that RPI4 and Jetson Nano platforms can consume only 56.06 kWh and 87.6 kWh per year, which can cost 1.81 USD and 2.8 USD, respectively. On the other hand, the adopted data fusion based energy efficiency system can save more than 20\% of the power consumption (as reported in Section \ref{sec1}), which represents more than 14726.8 kWh per year and hence a cost of more than 471.25 USD can be saved. In addition, if we take into consideration that an RPI4 costs 59.95 USD and Jetson Nano costs 99 USD, it can be deduced that using a RPI4, 411.3 USD can be saved while 372.25 USD can preserved if the Jetson Nano is deployed. Finally, it is worth mentioning that the new strategy to save energy that we have proposed can consume a limited amount of power in comparison with the large amount of energy that can be saved.

\begin{table} [htbp]
\caption{Yearly power consumption of a typical household in Qatar and the RPI4 and Jetson Nano platforms.}
\label{PC-Comp}
\begin{center}

\begin{tabular}{lccc}
\hline
& {\small \textbf{Per household (5.5 person)}} & {\small \textbf{Raspberry
Pi4} \ \ \ \ \ \ \ } & {\small \textbf{Jetson Nano} \ \ \ \ \ \ \ \ \ \ \ }
\\ \hline
{\small Consumption per year (kWh)} & \multicolumn{1}{l}{\small 73634} & 
\multicolumn{1}{l}{\small 56.06} & \multicolumn{1}{l}{\small 87.6} \\ 
{\small Cost per year (US Dollar)} & \multicolumn{1}{l}{\small 2356.28} & 
\multicolumn{1}{l}{\small 1.81} & \multicolumn{1}{l}{\small 2.8} \\ 
{\small Saved consumption (kWh)} & \multicolumn{1}{l}{{$>$ {\small 14726.8}}}
& \multicolumn{1}{l}{\small -} & \multicolumn{1}{l}{\small -} \\ 
{\small Saved cost (US Dollar)} & \multicolumn{1}{l}{{$>$ {\small \textbf{%
471.25}}}} & \multicolumn{1}{l}{{{\small \textbf{411.3}}}} & 
\multicolumn{1}{l}{{{\small \textbf{372.25}}}} \\ \hline
\end{tabular}

\end{center}
\end{table}

\section{Open challenges and future research orientations} \label{sec4}

Based on the literature review, data fusion taxonomy and discussion conducted in the previous sections, we further point out a set of of open challenges and recommend some prospect research orientations.

\subsection{Open challenges}
\subsubsection{Sensor selection}
Selecting sensors is still a significant challenge in energy efficiency and data fusion because the efficacy of the latter is impacted by the quality of the opted sensors, their number and their reliability. To implement a powerful energy efficiency system, a vast number of sensors is required, which produces several information formats \cite{Chakraborty2019}. Moreover, some sensors can be regarded as less significant than others while some are notably unworthy. Furthermore, data are generated with distinct sampling frequencies, and hence resulting in a serious variation of data resolution, accuracy and trustworthiness \cite{Guo7978501}.   

In addition, recording, handling and managing data from various modalities induce high computing and large communication costs, making the task of supporting real-time applications very hard. Therefore, a source selection stage should be set for judging the reliability of
information sources. This can be a wise decision that can decrease the computational complexity and communication overhead, especially at the gateway, in which data are aggregated and fused. In this regard, three principal challenges should be investigated as follows:
\begin{itemize}
\item Developing source-selection algorithms using internal and external confidence levels for evaluating the trustworthiness of data sources and via adopting deep learning models \cite{XU2017410,Meiseles8949507,Bascol8803325}. The latter represents a promising solution to develop efficient source-selection models.
\item Evaluating data utility through measuring the quality of collected information from the chosen sensors and reducing the energy consumed during the wireless communication process. This can be done for example by considering the spatial-temporal correlation and link quality of collected data \cite{Bijarbooneh7332242,Fei2019Sensors,Tayeh8689010}. 
\item Installing appropriate sensors and adequately selecting enough sensors to optimize energy efficiency in buildings.  
\end{itemize}

\subsubsection{Security and privacy preservation}
Implementing data fusion using various sources of data can result in revealing some private information, including consumer's location, preferences and daily activities, etc. In this regard, data fusion without considering privacy preservation and security issues can considerably affect user's acceptance of an energy efficiency solution. Therefore, privacy preserving and data transmission security are considered to be crucial requirements for data fusion in energy efficiency \cite{GUNDUZ2020107094}.  
To that end, the privacy preservation and security issue of data fusion is always an open research topic. Although privacy-friendly sensors can be used to preserve user privacy, they can not supply adequate information to satisfy application development purposes. How to warrant a trade-off between privacy preservation and providing enough data for energy efficiency ecosystems is still an open issue that requires more research investigations. In data fusion based energy efficiency frameworks, two potential solutions can be identified to ensure the privacy:
\begin{itemize}
\item The fusion of sensitive information (e.g. region/location, building ID, user ID, contact information, etc.) into generic data before transmitting them to the computing platforms for further processing. 
\item The use of generative deep learning models such as generative adversarial networks (GANs) to produce synthetic databases with similar data features for analysis and study objectives \cite{Esteban2017,Jones2019}. This helps in promoting data sharing after eliminating the chance that the private information will be leaked out. This is a promising solution that should be thoroughly investigated to set an explicit line between generic and sensitive data.
\end{itemize}   
On the other hand, data security is guaranteed using encryption methods, in which data are encoded in order that only the authorized parties can access them. This enables a scalable key management, in-depth access monitoring and flexible data sharing. Moreover, encryption can also be deployed to secure data transmission  between IoT sensors and computing platforms to avoid data hijacking \cite{Elhoseny2018}.

\subsubsection{Platform architecture}
The rise of cloud computing have enabled singular access to data through presumably secure servers on the internet. Many cloud services offer real-time (or almost real-time) performance. Also, they may provide a package of additional useful features to end-users to facilitate data processing and analysis. Thus, in the grand scheme, a cloud infrastructure is on the rise to become the default option for data management for different systems, including energy efficiency systems \cite{MARTINLOPO2020107101,UYSAL20171387}. 

On the flip side, issues of privacy and cybersecurity arise, especially in cloud-based solutions. Questions of whether the data is completely safe at a third-party server can become serious. Also, even if the data is stored in a cloud server managed by the data owner, questions on the security of the system may come to light. For instance, how well the system is protected against cybersecurity attacks? This enquiry is of crucial importance, especially when dealing with detailed energy consumption profiles that can be exploited in an adversely negative manner. This is one of the core motivations behind relying on local management of data completely isolated from the internet and only accessible to end-users in the intranet of the building (e.g. domestic home, school, business) \cite{ARI2019}. Protection is almost guaranteed in terms of data privacy and from external attacks (except in situations where physical intrusion is present or the intranet is hacked). All in all, the local storage solution can be considered quite plausible when no external connection to the internet is necessary. 
However, a system completely isolated from the internet can swiftly become outdated due to the constant change of end-users' habits, behavior patterns, and corresponding analysis of such changes. Unless the nature of data is fairly non-changing, retraining the system's algorithms with new external data is necessary. Inevitably, a hybrid solution will arise, creating a balance between local storage and cloud management. Sensitive data can be stored on-premises on a private cloud server. For public data, system algorithms and user interfaces can be managed on a cloud server on the internet \cite{FORCAN2020101988}.

Moreover, balancing the platform architecture cost and computing charge for information fusion is also a remaining issue, which depends mainly on the specific application scenario. In addition, developing flexible and modular platform architectures is still of utmost importance, which aims at (i) integrating IoT-based smart-sensors and actuators with innovative energy services; (ii) allowing the integration of heterogeneous hardware/software components \cite{IoTbook2018}; and (iii) permitting the implementation and execution of multi-modal sensor networks \cite{Aranda2020}.

\subsection{Future orientations}

\subsubsection{Deep learning for data fusion}
As an essential orientation, exploring more application possibilities for ML based information fusion is a promising research direction. With the major boost and advance of acceleration technologies such as the graphics processing unit (GPU) cards and cloud platforms, doors have been opened to build advanced data fusion models based on new ML techniques in order to generate highly accurate fusion results \cite{CHOI2019259}.  

To that end, the deployment of high complex and large-scale learning schemes in information fusion is of utmost importance. Much expectation has been placed on deep learning that relies on combining supervised and unsupervised training to create a hierarchization of learning, called the network. 
Particularly in some cases related to big data analysis, deep learning can provide a higher fusion fidelity and forecasting accuracy than other conventional ML approaches \cite{ZHANG2018146}. More specifically, some powerful solutions have been already proposed to construct data fusion frameworks using deep learning, as explained in \cite{Furqan2017}. 
In \cite{Vielzeuf8516399}, an interesting data fusion strategy is presented using deep learning that aims at effectively and systematically regulating the weights of low and high-level data fusion. In \cite{LIU2020123}, recent data fusion techniques based on deep learning used in urban traffic are briefly reviewed. Nonetheless, other subsequent challenges have appeared at the same time. On the flip side, we expect also that data fusion mechanisms can be deployed to increase the efficiency of deep learning to deal with large-scale and real-time data from heterogeneous sources. This will help in training all the data together as mentioned in \cite{GUO2019215}. 

Besides, it is known that the efficacy of deep learning is essentially guaranteed with significant amount of data and large resource consumption. How to guarantee perfect application of deep learning into data fusion based energy efficiency ecosystems with low computational devices and small amount of data is another challenge. In addition, ensuring a good compromise between fusion effectiveness and quality is an extra issue to be resolved in the future.
Excepting the concerns mentioned above, researchers are strongly invited to contribute in developing of deep composite intelligent energy efficiency frameworks. \vskip 4mm

\subsubsection{Cryptocurrencies and blockchain for energy efficiency}
Nowadays, application of cryptocurrencies including Bitcoin and other technologies such as blockchain have emerged and affected other research areas such as data management, healthcare, economy, smart city, etc. Similarly, these tools present both interesting perspectives and potentials to the energy efficiency sector. Specifically, cryptocurrencies can be deployed to persuade consumers take energy saving actions. In that respect, novel green cryptocurrencies can also be created and used to reward end-users according to their feedback and energy efficiency behavioral change \cite{JOACHAIN201489}.

\subsubsection{Explainable recommender systems}
Data fusion is desired to design a powerful energy efficiency system through (a) performing an adequate analysis of large-scale data from multiple sources; and (b) providing personalized recommendations for end-users to reduce energy wastage. Towards this end, increasing users' acceptance rates of these recommendations is of great importance. Explainable recommendations is promising solution attempting to build models that can not only trigger personalized high-level recommendations but provide intuitive explanations as well \cite{Zhang2018ERS,REHABC2020}. The explanations can be post hoc or even promptly coming from an explainable system that should be developed based on data fusion analysis. Explainable recommender systems aim at addressing the question of why? via producing explanations to consumers or building owners to allow them comprehend why a particular energy saving action is recommended by the recommendation model.

\subsubsection{Novel visualization tools}
As discussed in Section \ref{sec2}, several criteria could strongly influence the quantity and quality of the consumed energy at all types of buildings. In pace with the current advancement of technology, data visualization and comprehension of energy feedback are key factors that play a major role in the extent to which visualization tools are likely to result in endorsing behavioral changes for energy efficiency \cite{CARNEIRO2019330}. Moreover, visualization plots are not only used to represent the data involved in the process in so many different ways, but they are also utilized as a data fusion strategy. To this end, investigating novel data visualizations is a high-demand topic, which aims at designing inventive, interactive and real-time tools that can intuitively facilitate the comprehension of consumption fingerprints \cite{GARCIA201895}.

\subsubsection{Fusion of 2D descriptions}
As illustrated in Section \ref{sec3}, the use of 2D local descriptors for appliance identification and NILM applications is a promising research area. A large variety of 2D descriptors have been proposed and successfully applied in various research topics. The energy efficiency community can benefit from those experiences to build powerful NILM systems \cite{Himeur2020IntelliSys}. In this context, the data fusion of 2D descriptors plays a major role in developing such solutions that not only improves the appliance identification at almost no cost, but also has a positive impact on promoting energy efficiency in buildings via providing the end-user with appliance-level consumption fingerprints.

\section{Conclusion} \label{sec5}
In this article, recent developments in data fusion for the energy efficiency applications have been reviewed and their contribution in implementing different scenarios of energy efficiency ecosystems has been described. We discussed the usefulness of several data fusion strategies that have been implemented or could be deployed to decrease energy wastage and promote sustainability. We also presented a generic taxonomy of energy efficiency frameworks involving data fusion schemes. Moreover, a comparison between various data fusion based energy efficiency frameworks has been conducted based on the aforementioned perspectives. On the other side, to illustrate the significance of data fusion in energy efficiency, a case study scenario has been presented, in which a novel appliance identification approach based on the fusion of 2D descriptors is proposed.

Moving forward, on the basis of the conducted survey and taxonomy, we came with important conclusions on how to improve the quality of data fusion based energy efficiency systems and reduce their costs. In this regard, an ensemble of remaining open issues has been identified, which can be summarized as follows:
\begin{itemize}
\item Adopting deep learning models to implement efficient source-selection models based on internal and external confidence levels in order to assess the trustworthiness of data sources.
\item Endorsing the use of novel strategies to preserve security and privacy via the fusion of sensitive information (e.g. region/location, building ID, user ID, contact information, etc.) into generic data. Moreover, the use of generative deep learning models, including GANs to produce synthetic databases  can help to promote data sharing while ensuring confidentiality and information privacy.
\item Balancing platform architecture costs and computing charges when implementing fusion strategies, and further developing flexible and modular platform architectures to implement effective energy saving ecosystems.
\item Using hardware platforms enabling more cost-effective and powerful alternatives to process and transmit metered power consumption and contextual information. This is a high priority because the main objective of energy efficiency is to reduce energy costs.
\end{itemize}
On the other hand, several future recommendations and directions have been derived to improve the performance of data fusion based energy efficiency systems. Specially, a set of challenging directions has been presented and explained, among them:
\begin{itemize}
\item The use of deep learning to build  powerful data fusion strategies ensuring a higher fusion fidelity and forecasting accuracy than other existing techniques.
\item The application of  emerging technologies, such as cryptocurrencies (e.g. Bitcoin) and blockchain to endorse an energy saving behavior among end-users. For example, cryptocurrencies have an essential role when they are used to motivate end-users behavioral change towards the adoption of an energy saving lifestyle.
\item The deployment of explainable recommender systems aiming at providing end-users with personalized advices to reduce wasted energy, followed by explanations about them.
\item The implementation of innovative visualization tools, in which the benefits are two-fold. First, they can be utilized as a fusion strategy to visualize the relation between energy consumption and contextual information. Second, they are deployed to understand energy consumption footprints and hence endorse behavioral changes.
\item The fusion of 2D local descriptors to extract appliance-level consumption footprints from aggregated consumption. This solution is very promising since 2D descriptors allow to extract fine-grained appliance features and their performance could be improved tremendously using fusion strategies.
\end{itemize} 
Finally, we are certain that this framework will be a concise and comprehensive reference for researchers and practitioners in the field of data fusion based energy efficiency systems. Additionally, an in-depth analysis could be more expanded into specific applications to precisely investigate the exhaustive requirements and deployed methodologies, which are not included in this framework.

\section*{Acknowledgements}\label{acknowledgements}
This paper was made possible by National Priorities Research Program (NPRP) grant No. 10-0130-170288 from the Qatar National Research Fund (a member of Qatar Foundation). The statements made herein are solely the responsibility of the authors.

\section*{Appendix }\label{}
Table \ref{AbbDescription} provides the acronym definitions of the energy efficiency projects described in Section 2.9.
 
\begin{table}[htbp]
\caption{Acronym definitions of the energy efficiency frameworks discussed in Section \ref{EngFrameworks} .}
\label{AbbDescription}
\begin{center}

\begin{tabular}{lc}
\hline
\textbf{Acronym} & \textbf{Description} \\ \hline
{\small BIGFUSE} & \multicolumn{1}{l}{\small Semantics for Big Data Fusion
and Analysis: Improving energy efficiency in smart grids} \\ 
{\small STREAMER} & \multicolumn{1}{l}{\small Semantics-driven Design
through Geo and Building Information Modelling for Energy-efficient } \\ 
& \multicolumn{1}{l}{\small Buildings Integrated in Mixed-use Healthcare
Districts} \\ 
{\small BENEFFICE} & \multicolumn{1}{l}{\small BENEFFICE: Behaviour Change,
Consumption Monitoring and Analytics with } \\ 
& \multicolumn{1}{l}{\small Complementary Currency Rewards} \\ 
{\small ORBEET} & \multicolumn{1}{l}{\small ORganizational Behaviour
improvement for Energy Efficient administrative public offices} \\ 
{\small MOBISTYLE} & \multicolumn{1}{l}{\small MOtivating end-users
Behavioral change by combined ICT based tools and modular} \\ 
& \multicolumn{1}{l}{\small \ Information services on energy use, indoor
environment, health and lifestyle} \\ 
{\small ChArGED} & \multicolumn{1}{l}{\small CleAnweb Gamified Energy
Disaggregation} \\ 
{\small GreenPlay} & \multicolumn{1}{l}{\small Game to promote energy
efficiency actions} \\ 
{\small TRIBE} & \multicolumn{1}{l}{\small TRaIning Behaviours towards
Energy efficiency: Play it!} \\ 
{\small EnerGAware} & \multicolumn{1}{l}{\small Energy Game for Awareness of
energy efficiency in social housing communities} \\ 
{\small PEAKapp} & \multicolumn{1}{l}{\small Personal Energy Administration
Kiosk application: an ICT-ecosystem for Energy  } \\ 
& \multicolumn{1}{l}{\small Savings through Behavioural Change, Flexible
Tariffs and Fun} \\ 
{\small enCOMPASS} & \multicolumn{1}{l}{\small Collaborative Recommendations
and Adaptive Control for Personalised Energy Saving} \\ 
{\small ENTROPY} & \multicolumn{1}{l}{\small Design of an innovative
energy-aware it ecosystem for motivating behavioural changes} \\ 
& \multicolumn{1}{l}{\small towards the adoption of energy efficient
lifestyles} \\ 
{\small GAIA} & \multicolumn{1}{l}{\small Green Awareness in Action} \\ 
{\small GreenSoul} & \multicolumn{1}{l}{\small Eco-aware Persuasive
Networked Data Devices for User Engagement in Energy Efficiency} \\ 
{\small BESTEnergy} & \multicolumn{1}{l}{\small Built Environment
Sustainability and Technology in Energy} \\ 
{\small eSESH} & \multicolumn{1}{l}{\small Saving Energy in Social Housing
with ICT} \\ 
{\small SmartBuild} & \multicolumn{1}{l}{\small Implementing Smart
Information and Communication Technology (ICT) concepts} \\ 
& \multicolumn{1}{l}{\small for energy efficiency in public buildings} \\ 
{\small SAVE@WORK} & \multicolumn{1}{l}{\small The Energy Saving Contest for
Public Authorities} \\ 
{\small EM3} & \multicolumn{1}{l}{\small Consumer Engagement Towards Energy
Saving Behavior by means of Exploiting} \\ 
& \multicolumn{1}{l}{\small Micro Moments and Mobile Recommendation Systems}
\\ \hline
\end{tabular}

\end{center}
\end{table}


\end{document}